\documentclass[11pt,letterpaper]{article}
\usepackage[margin=1in]{geometry}
\usepackage{amsmath,amssymb}
\usepackage{booktabs}
\usepackage{graphicx}
\usepackage{setspace}
\usepackage{natbib}
\usepackage[hidelinks,hypertexnames=false]{hyperref}
\hypersetup{pdftitle={Whose Households? Trade, Investment, and the
Rise of Services},
pdfauthor={Li Gan}}
\usepackage{threeparttable}
\usepackage{float}

\newtheorem{prop}{Proposition}

\newcommand{\RNfellAll}{61}
\newcommand{\RNmedDRothers}{8.6}
\newcommand{\RNslopeDRDs}{0.17}
\newcommand{\RNslopeDRDsSE}{0.22}

\newcommand{\RNmedRninetyfive}{0.553}
\newcommand{\RNmedRoeight}{0.513}
\newcommand{\RNmedReighteen}{0.495}
\newcommand{\RNshareByOeight}{69}
\newcommand{\RNcommonA}{0.059}
\newcommand{\RNcommonAse}{0.026}
\newcommand{\RNcommonB}{0.007}
\newcommand{\RNcommonBse}{0.015}
\newcommand{\RNincSlope}{0.108}
\newcommand{\RNincSlopeSE}{0.046}
\newcommand{\RNmeanDlogR}{0.129}

\newcommand{\RNincomePart}{0.063}
\newcommand{\RNincShare}{0.49}
\newcommand{\RNincShareLo}{0.08}
\newcommand{\RNincShareHi}{0.90}
\newcommand{\RNwithinR}{0.088}
\newcommand{\RNwithinRse}{0.034}
\newcommand{\RNwithinRnoAsia}{0.077}
\newcommand{\RNwithinRnoAsiaSE}{0.054}
\newcommand{\RNbetweenR}{0.221}
\newcommand{\RNbetweenRse}{0.045}
\newcommand{\RNbetweenRpop}{0.148}
\newcommand{\RNbetweenRpopSE}{0.036}
\newcommand{\RNinvDom}{-0.9}
\newcommand{\RNinvDomRose}{37}
\newcommand{\RNinvFrn}{+2.2}
\newcommand{\RNinvFrnRose}{86}

\newcommand{\RNhhAnyLateFell}{23}
\newcommand{\RNrisersLate}{61}
\newcommand{\RNpcDom}{0.05}
\newcommand{\RNpcGlob}{0.42}
\newcommand{\RNpcGlobMin}{0.21}
\newcommand{\RNpcGlobMax}{0.43}
\newcommand{\RNpcDomMin}{0.02}
\newcommand{\RNpcDomMax}{0.14}
\newcommand{\RNcovA}{0.97}
\newcommand{\RNcovSig}{0.16}
\newcommand{\RNcovS}{0.18}
\newcommand{\RNcovSA}{0.0005}
\newcommand{\RNcovSSig}{0.0355}
\newcommand{\RNvarS}{0.0553}
\newcommand{\RNlevSlope}{0.35}
\newcommand{\RNlevSlopeSE}{0.17}
\newcommand{\RNiqrLo}{0.42}
\newcommand{\RNiqrHi}{0.58}
\newcommand{\RNiqrAnyLo}{0.65}
\newcommand{\RNiqrAnyHi}{0.74}
\newcommand{\RNHlcuN}{157}
\newcommand{\RNHusdMed}{0.45}
\newcommand{\RNHlcuMed}{0.47}
\newcommand{\RNHusdMaj}{41}
\newcommand{\RNHlcuMaj}{44}
\newcommand{\RNHflips}{7}
\newcommand{\RNexpLcuMed}{33}

\newcommand{\RNHgradLcu}{0.101}
\newcommand{\RNHgradLcuSE}{0.017}
\newcommand{\RNHgradRone}{0.016}
\newcommand{\RNHgradRoneSE}{0.014}
\newcommand{\RNexpShare}{38}
\newcommand{\RNexpMaj}{31}
\newcommand{\RNHtabPoolMed}{0.474}
\newcommand{\RNHtabPoolMaj}{0.44}

\newcommand{\RNHtabWaMed}{0.496}
\newcommand{\RNHtabWaMaj}{0.48}
\newcommand{\RNHtabWaN}{58}
\newcommand{\RNHtabWbMed}{0.486}
\newcommand{\RNHtabWbMaj}{0.47}
\newcommand{\RNHtabWbN}{62}
\newcommand{\RNHtabWcMed}{0.394}
\newcommand{\RNHtabWcMaj}{0.32}
\newcommand{\RNHtabWcN}{37}
\newcommand{\RNHtabTloMed}{0.587}
\newcommand{\RNHtabTloMaj}{0.77}
\newcommand{\RNHtabTloN}{53}
\newcommand{\RNHtabTmidMed}{0.439}
\newcommand{\RNHtabTmidMaj}{0.31}
\newcommand{\RNHtabTmidN}{52}
\newcommand{\RNHtabThiMed}{0.376}
\newcommand{\RNHtabThiMaj}{0.23}
\newcommand{\RNHtabThiN}{52}
\newcommand{\RNexpVolShare}{39}
\newcommand{\RNdetN}{1{,}800}
\newcommand{\RNdetG}{75}
\newcommand{\RNdetBN}{75}

\newcommand{\RNdetWAabs}{0.164}

\newcommand{\RNdetBAabs}{0.265}

\newcommand{\RNdetWSigabs}{0.111}

\newcommand{\RNdetBSigabs}{0.169}

\newcommand{\RNdetWSabs}{0.036}
\newcommand{\RNdetWSSE}{0.028}
\newcommand{\RNdetBSabs}{0.125}

\newcommand{\RNdetWRabs}{0.088}

\newcommand{\RNdetBRabs}{0.221}
\newcommand{\RNdetWshCabs}{0.0797}
\newcommand{\RNdetBshCabs}{0.0931}

\newcommand{\RNdetBshGabs}{0.0291}
\newcommand{\RNdetWshIabs}{0.0739}
\newcommand{\RNdetBshIabs}{0.0133}

\newcommand{\RNdetBshXabs}{0.0773}

\newcommand{\RNpbShareFiftysix}{48}
\newcommand{\RNallCompFiftysix}{55}
\newcommand{\RNcorrDsCpVol}{0.32}
\newcommand{\RNcorrDRCpVol}{0.81}

\newcommand{\RNvolShareByOeight}{75}
\newcommand{\RNvolCommonB}{0.030}
\newcommand{\RNvolCommonBse}{0.014}
\newcommand{\RNvolIncShare}{0.38}
\newcommand{\RNvolIncSlope}{0.108}
\newcommand{\RNvolIncSlopeSE}{0.054}
\newcommand{\RNgradAnyW}{0.058}
\newcommand{\RNgradAnyWse}{0.015}
\newcommand{\RNgradDomW}{0.054}
\newcommand{\RNgradDomWse}{0.019}
\newcommand{\RNgradAnyB}{0.041}
\newcommand{\RNgradAnyBse}{0.009}
\newcommand{\RNgradDomB}{0.093}
\newcommand{\RNgradInvAnyW}{0.062}
\newcommand{\RNgradInvAnyWse}{0.013}
\newcommand{\RNgradInvDomW}{0.052}
\newcommand{\RNgradInvDomWse}{0.015}
\newcommand{\RNgradDomBse}{0.015}
\newcommand{\RNanyFellA}{49}

\newcommand{\RNfrnRatioC}{1.03}

\title{\large Whose Households?\\
Trade, Investment, and the Rise of
Services\thanks{The country-year and episode datasets accompany the
paper. The construction and analysis code is available from the author
and will be deposited in a permanent public archive upon publication.
AI software (Anthropic's Claude) was used under the author's direction
to write and check the data-processing and accounting code, check the
algebra and the literature, and edit the draft. All output was checked
by the author, who is responsible for the content. The author declares
no competing interests.}}
\author{Li Gan\thanks{Department of Economics, Texas A\&M University,  Email:
\texttt{ganli@tamu.edu}.}}
\date{October 2026}

\begin{document}
\maketitle

\begin{abstract}
\noindent As economies become service economies, whose demand do their
service sectors serve? I trace service value added through harmonized
input-output tables for 76 economies over 1995--2022. The tracing runs
first to domestic final demand and then, with inter-country tables, to
final users anywhere in the world, at current prices and in chain
volume. Five findings follow. First, the service share does not reveal
how much of a country's services its own households support: economies
with nearly the same share differ by 21 to 39 percentage points. Second,
that domestic household share fell in 61 of the 76 economies, including
50 of the 65 whose service share rose, and most of the fall came before
2008. Third, the support moved to two places. Exports moved it
abroad, about equally to foreign households and to foreign investment.
When foreign households are added to domestic ones, the fall is half as
large. Households at home and abroad still support most of the services
of 75 of the 76 economies. Within countries, support moved from
households to investment as income grew.
Fourth, whose demand supports services matters. Investment and exports
support a median 39 percent of service value added but carry 59 percent of the
variance of its real growth. Household-supported service value added is
the steadiest component in 71 of 76 economies. The service share also
understates the income effect in household service demand: it rises
less with income than the service content of household demand does.
Fifth, an exact accounting allocates most of the growth in the service
share to the production system: the production coefficients and the
industry composition of investment and export demand. Almost all of the
fall in the domestic
household share goes to the final users' shares in total final demand,
not to what households buy.
\medskip

\end{abstract}

\section{Introduction}\label{sec:intro}

As economies grow richer, more of what they produce is services. The
standard reading is that their households are buying more services
\citep{buera2012rise,herrendorf2014growth}. For the domestic economy,
this paper shows, the reading fails. Since 1995 the part of each
country's services that its own households support has fallen almost
everywhere. It fell in 61 of 76 economies, and in 50 of the 65 whose
service share rose. Most of the fall came before 2008. Two movements
account for it. Exports moved support abroad, about equally to foreign
households and to foreign investment. Within countries, as income
grew, support moved from households to investment. Households, at home
or abroad, remain the main final users of services almost everywhere.
But less of the support for each
country's services comes from its own households, and that changes how
its services move with the cycle.

The service share records where value added is produced, not whose
final demand sustains it. A catered business meal, a logistics
contract for an exporter, and a family's restaurant meal enter the
share in the same way. When a manufacturer's accounting office becomes
a separate business-services firm, measured tertiarization jumps
although nothing changes in whom the economy serves. In
\citet{buera2012rise}, services are final consumption goods, so every
unit of service value added is supported by domestic households by
construction. But investment and exports can also raise the service
share, through outsourcing and supply chains that reach service
producers \citep{berlingieri2014outsourcing,garciasantana2021investment,
gaggl2026structural}.

I measure the purchaser side directly. Using harmonized input-output
tables for 76 economies over 1995--2022, I trace each service
industry's value added through domestic supply chains to the final
demand that sustains it. There are four final users: domestic
households, government, investment, and exports. The
\textit{domestic household share} of the service economy is
$R=V_C^S/V^S$, the part $V_C^S$ of market-service value added $V^S$
that domestic household consumption supports. I then follow exports across borders with inter-country
tables, which tells whether a foreign household, a foreign government,
or a foreign investor finally absorbs them. The main results are
reported at current prices and in chain volume, using the OECD's
tables in previous-year prices.

The paper makes three contributions. First, a statistic. The domestic
household share is the part of a country's service value added that its
own households support; a counterpart counts households anywhere. Both
are built for 76 economies over 1995--2022 and released with the paper.
What value-added trade accounting did for gross exports
\citep{johnson2012accounting,johnson2017portrait}, this does for the
service share: it says whose demand the number reflects. Second, a
fact. By this measure the rise of services since 1995 was not carried
by domestic households, almost anywhere, and most of the shift came in
the trade-integration years. The standard reading of the service share
fails for the domestic economy. Third, consequences. The demand behind
services sets how they move with the cycle, and the service share
understates how household service demand rises with income.

Five findings follow. First, the service share does not reveal whose
demand supports services. Economies with nearly the same service share
differ by 21 to 39 percentage points in $R$. China and India have
almost the same service share in 2018 (0.41 and 0.40), yet domestic
households support 42 percent of China's market-service value added
and 58 percent of India's. Across economies, the service share shows
no detectable relation to $R$ once income is controlled for. It does carry
information about the share that households anywhere, at home or
abroad, support (partial correlation \RNpcGlob{} in 2018). The
difference between the two is the trade route.

Second, the domestic household share fell, almost everywhere and
mostly before 2008. Among the 65 economies whose service share rose over
1995--2018, $R$ fell in 50 (77 percent); in chain volume, in 54 of 64.
But $R$ also fell in all 11 economies whose service share did not rise.
And \RNshareByOeight{} percent of the fall in the median came before 2008
(\RNvolShareByOeight{} percent in chain volume). Economies with faster
income growth saw larger falls. The exact identity
\[
R=\frac{\alpha_C\,\sigma_C}{s^S}
\]
shows the mechanics. Here $\alpha_C$ is the household share of value
added, the part of all value added that household consumption supports.
$\sigma_C$ is the service content of household demand, and $s^S$ is the
service share. As income rises, $\sigma_C$ rises, the familiar Engel
pattern, but $\alpha_C$ falls faster.

Third, the support moved to two different places. Over 1995--2008,
exports moved it abroad. Among
economies with rising service shares, the median domestic household
share fell 4.5 points, while foreign households gained 2.2 points and
foreign investors, reached through exports, 2.1. Foreign households
absorb two-thirds of export-supported service value added. So counting
households anywhere halves the fall over 1995--2018. Households at home
and abroad still support most of the services of 75 of the 76
economies. China is the one exception, at 49 percent, because domestic and foreign
investment support 37 percent of its services. Within countries, support
moved toward investment instead as income grew: the share of services
that households anywhere support falls with income as much as the
domestic household share does.

Fourth, whose demand supports services matters. Services inherit the
volatility of the demand behind them. Investment and exports support a
median 39 percent of market-service value added but carry 59 percent of
the variance of its real growth. Household-supported service value
added is the steadiest component in 71 of 76 economies. So part of why
services are the stable sector is also whom they serve. The demand
mix of a sector is the share of its value added that each final user
supports. In the median economy, giving goods the demand mix of
services closes 37 percent of the volatility gap between the two
sectors. In 2009, household-supported
service value added grew in the median economy while export- and
investment-supported value added fell. Whose demand supports services
also matters for inference. A model that equates the service share with
household demand assigns household service demand a weaker income
effect than the service content of household demand shows. Within
countries, net of common year effects, the two income gradients differ
by 0.14 (s.e.\ 0.06).

Fifth, an exact accounting of the 1995--2018 changes allocates most of
the growth in the service share to the production system. The median
share is 73 percent (67 percent in chain volume). Of it, the production
block alone, the input and value-added coefficients, receives 48
percent; the rest is the industry composition of investment and export
demand. Almost all of the fall in $R$ goes to the final-demand
structure, the shares of households, government, investment, and
exports in total final demand. The composition of household demand,
where a change in what households buy would show, raised $R$ where it
moved and was overwhelmed.

The paper builds on work that splits structural change by expenditure
category. \citet{herrendorf2013two} show that measuring sectors by
final expenditure or by value added changes the inferred roles of
income and prices. They build United States consumption value added,
of which $\sigma_C$ is the household-only counterpart.
\citet{garciasantana2021investment} write a sector's value-added share
as the investment, consumption, and export weights times the sector's
share within each. They show that investment drives part of structural
change. $R$ is the consumption term of that decomposition, restricted
to households and expressed as a share of services.
\citet{herrendorf2021investment} show that investment has its own
sectoral composition. \citet{gaggl2026structural} document for the
United States that services supply a rising share of intermediates and
of new capital. \citet{casal2024investment} show that investment
networks change with development and matter for cross-country income
differences. Here domestic households are separated from government,
and the export term is followed abroad by type of final user.

Value-added trade accounting traces value added to the country and
product where it is finally absorbed \citep{johnson2012accounting,
johnson2017portrait,timmer2014slicing,los2015global}. The OECD
publishes domestic value added in foreign final demand and the
value-added content of consumption by source industry
\citep{oecd2023tiva}. The rise of export-supported service value added
in 1995--2008 documented here is the service side of the expansion that
literature measures. A
separate literature shows that trade shapes sectoral composition
\citep{matsuyama2009structural,uy2013structural,
kehoe2018global,lewis2022structural,sposi2026trade}. Another links the
sectoral composition of output to aggregate volatility through sector
productivity shocks \citep{moro2012structural,moro2015structural,
carvalho2013great}. Section~\ref{sec:exposure} adds a demand-side
channel.

Industry classifications answer a related question.
\citet{fan2023india} split consumer from producer services by whether
firms sell to consumers or to other firms.
\citet{chen2023tertiarization} distinguish consumer-service from
producer-service industries. \citet{duernecker2024services} and
\citet{buera2022skill} split services by productivity growth and skill
intensity. A firm's direct buyer does not fix the final one. In 2015,
firms directly bought 41 percent of China's accommodation-and-food
output in the harmonized tables, against 15 percent at the world
median (Appendix Table~\ref{tab:sectors}). Appendix~\ref{app:validation}
checks the cell against the official tables. Yet some of those
purchases ultimately serve households. So I
classify service value added by the final demand that sustains it.

Section~\ref{sec:framework} defines the measures.
Section~\ref{sec:twotypes} shows that the service share does not
reveal them, documents the fall in $R$, and separates its two sources.
Section~\ref{sec:global} follows exports to their final users.
Section~\ref{sec:determinants} decomposes $R$ exactly,
Section~\ref{sec:accounting} allocates the observed changes, and
Section~\ref{sec:prices} reports chain-volume results.
Section~\ref{sec:inference} shows why whose demand supports services
matters: for how service output moves, and for what the service share
says about household demand. Section~\ref{sec:conclusion} concludes.

\section{Framework and Measurement}\label{sec:framework}

\subsection{Data}

I use the OECD's 2025 release of harmonized national input-output
tables, observed annually over 1995--2022 with 50 ISIC Rev.\ 4
industries. The release covers 80 economies; the analysis uses a fixed
sample of 76 economies, yielding 2,128 country-year tables. The data
include domestic use tables with import matrices and value-added and
gross-output rows. Two further OECD sources extend the analysis. The
inter-country input-output tables (2026 release; 80 economies and the
rest of the world, the same 50 industries) let me follow exports to
their final users (Section~\ref{sec:global}). Their companion tables
in previous-year prices give chain-volume versions of the main counts,
the timing of the fall, and the Shapley allocations (Section~\ref{sec:prices}). The current-price inter-country
tables reproduce the national-table measures: across the 2,128
country-years, the largest difference in $R$ is 0.0002.

One input-output operation, tracing, underlies every measure in the
paper. It follows
each unit of service value added past the immediate customer to the
final purchaser who ultimately sustains it. A trucking firm paid by a
manufacturer whose goods are exported is, in this accounting,
supported by export demand rather than by household consumption. The
tool is the Leontief inverse. The underlying production structure is
fixed-coefficients: producing one unit of industry $j$'s output
requires $A_{ij}$ units of industry $i$'s domestically produced output
as an intermediate input. So producing the gross-output vector $x$
absorbs intermediate inputs $A x$. Every unit of gross output is either
absorbed as an input or delivered to final demand $f$, giving the
material-balance identity and its solution
\[
x \;=\; A x + f
\qquad\Longrightarrow\qquad
x \;=\; Lf,
\qquad L\equiv(I_n-A)^{-1}.
\]
The Leontief inverse $L=I_n + A + A^2 + \cdots$ cumulates production
chains. Its $(i,j)$ entry is the output industry $i$ must produce,
directly and through every round of intermediate inputs, for one unit
of industry $j$'s output to reach final demand. The series converges
because the spectral radius of $A$ is below one in every table used. In
the current-price tables every column of $A$ sums to less than one; the
remainder is value added, imports, and net taxes. Multiplying the inverse by a
category's final-demand vector converts that category's purchases into
the gross output they ultimately call forth in every industry.

\textit{Market services} are the 17
service industries excluding public administration, education, and
health. The \textit{consumer-facing} subset (trade, accommodation and
food, recreation, personal services) is used where sector labels
matter. Full construction details, group definitions, and measurement
caveats are in Appendix~\ref{app:construction}.

\subsection{From direct purchasers to final demand}\label{sec:ultimate}

The two accounting layers answer different questions. The
\textit{direct} layer records who buys a service industry's output
immediately: intermediate users, households, government, investment,
or exports. The \textit{final-demand} layer follows intermediate
purchases through the domestic production network and assigns the
service activity to the final-demand category that sustains it. A
catered business meal is a direct sale to a firm. The final-demand
layer assigns it to whatever final demand that firm's output serves,
such as household consumption, investment, or exports. Let
$\mathcal K=\{C,G,I,X\}$ denote the four categories: household
consumption, government and NPISH consumption, investment (fixed
capital formation and changes in inventories), and exports. One rule
fixes the notation. A vector, with one entry per
industry, carries its category as a superscript. A value-added total
carries the supporting category as a subscript and the producing
sector as a superscript. So $f^k$ is the final-demand vector of
category $k$, and the gross output it supports is
\[
x^k=Lf^k.
\]
Let $v_i=\mathrm{VA}_i/x_i$ be industry $i$'s value-added coefficient and
$\widehat v=\operatorname{diag}(v)$; let $e_S$ select market-service
industries. The value added that category $k$ supports, in the whole
economy and in market services, is
\[
V_k=\mathbf{1}'\widehat v x^k,
\qquad
V_k^S=e_S'\widehat v x^k.
\]
Read $V_C^S$ as: supported by household consumption, the subscript;
produced in market services, the superscript. The assignments are
additive:
\[
V=\sum_{k\in\mathcal K}V_k,
\qquad
V^S=\sum_{k\in\mathcal K}V_k^S.
\]
Dropping the subscript means summing across all categories: $V$ is
total value added and $V^S$ is total value added produced in market
services.

The two headline measures use this assignment at two horizons:
\[
R_{ct}=\frac{V_{C,ct}^S}{V_{ct}^S},
\qquad
H_{ce}=\frac{\Delta V_{C,ce}^S}{\Delta V_{ce}^S}.
\]
Here $c$ indexes countries, $t$ years, and $e$ tertiarization episodes,
defined in Section~\ref{sec:classification}. $R$ is the \textit{domestic household share} of market-service value
added at a date. $H$ is the household contribution to the growth of
market-service value added over an episode. The two carry the same
information about the direction of change. The end-of-episode $R$ is a
weighted average of the starting $R$ and $H$, so $R$ rises exactly when
$H$ exceeds the starting $R$. I use $H$ only for one purpose that $R$
does not serve, which is to ask whether households supported most of
the new service value added. Both are accounting statistics: they say
whose demand supports services, not why that demand moved. The
direct layer is a diagnostic and is reported in
Appendix~\ref{app:direct}.

The tracing in this layer stops at the border. Exports are a final
category: a domestic service embodied in an exported consumer good is
export-supported even if a foreign household eventually buys the good.
Section~\ref{sec:global} removes that limit with inter-country tables.
It measures $R^{\mathrm{any}}$, the share of an economy's market-service
value added that households anywhere support, at home or abroad; $R$
counts only the economy's own households. The categories follow
national accounts, so residential construction is investment-supported
even when households choose the dwelling.
Government remains a separate category throughout, and
``non-household-supported'' is the exact residual label.
Appendix~\ref{app:definitions} shows that excluding government from the
scalar measure leaves the main comparisons unchanged.

The direct layer is informative because industry labels are imperfect
proxies for demand. At the world median, intermediate users buy 36
percent of consumer-facing service output. Intermediate use exceeds
household purchases of market services in 84 percent of economies
(Appendix Tables~\ref{tab:levels} and~\ref{tab:fractions}). Yet some of
those purchases ultimately serve households, such as
logistics used to deliver a retail sale. Tracing follows them. It can
only add to households' share of service gross output, and at the
median it adds 15 points (Appendix Figure~\ref{fig:ultimate} compares
the direct and traced shares). Even after tracing, domestic households
support only about half of market-service value added at the world
median: $R$ declines from 0.55 in 1995 to 0.50 in 2018.

\subsection{An exact identity for the domestic household share}

The level measure has an exact representation. Define
\[
\alpha_C=\frac{V_C}{V},\qquad
\sigma_C=\frac{V_C^S}{V_C},\qquad
s^S=\frac{V^S}{V},\qquad
R=\frac{V_C^S}{V^S}.
\]
Here $\alpha_C$ is the \textit{household share of value added}: the
part of all value added supported by household consumption. $\sigma_C$
is the \textit{service content of household demand}: the market-service
share of the value added that household consumption supports, directly
and through the supply chains of everything households buy. $s^S$ is
the service share. $\sigma_C$ is not the direct service-expenditure
share. It also moves with production linkages, value-added
coefficients, and relative prices, so it is consistent with, but not a
pure measure of, consumer upgrading. Then
\begin{equation}\label{eq:Ridentity}
\frac{\alpha_C \sigma_C}{s^S}
=
\frac{V_C}{V}\,
\frac{V_C^S}{V_C}\,
\frac{V}{V^S}
=
\frac{V_C^S}{V^S}
=R.
\end{equation}
The product $\alpha_C\sigma_C$ is household-supported service value
added as a share of all value added; dividing by $s^S$ expresses it as
a share of the service sector. The identity is accounting, not a
model. It is the consumption term of the decomposition in
\citet[eq.~1]{garciasantana2021investment}, restricted to households
and expressed as a share of the service sector. It shows why a rising
service content of household demand need not raise $R$: $R$ falls if
$\alpha_C$ falls or $s^S$ rises faster. A round-number illustration
shows the mechanics. If households support 40 percent of all value
added, 60 percent of that is market services, and market services are
half of value added, then $R=0.40\times0.60/0.50=0.48$. If $\sigma_C$
rises to 0.65 while $\alpha_C$ falls to 0.35, $R$ falls to 0.455,
although the service content of household demand has risen.
Section~\ref{sec:determinants} measures the three margins.

\section{Domestic Households in Levels and Changes}\label{sec:twotypes}

This section counts domestic households only.
Section~\ref{sec:reveal} shows that the service share does not reveal
$R$. Section~\ref{sec:trajectories} documents the fall in $R$ and
separates its two sources. Section~\ref{sec:classification} measures
the household contribution to service growth over tertiarization
episodes: country-windows in which both the market-service share $s^S$
and nominal market-service value added $V^S$ rise.

\subsection{Levels: what the service share does not
reveal}\label{sec:reveal}

The service share does not pin down $R$. Table~\ref{tab:canonical}
groups economies with similar service shares: within each band, $s^S$
differs by at most 2.3 points, while $R$ differs by 21 to 39 points. I
summarize each economy by the pair $(s^S, R)$: the market-service share
of aggregate value added and the domestic household share of that
service value added. Their cross-country relationship is not expected
to be simply positive. The identity $R = \alpha_C \sigma_C / s^S$
places $s^S$ in $R$'s denominator, so some negative covariance between
the two is arithmetic rather than evidence.

\begin{figure}[H]
\centering
\includegraphics[width=0.98\textwidth]{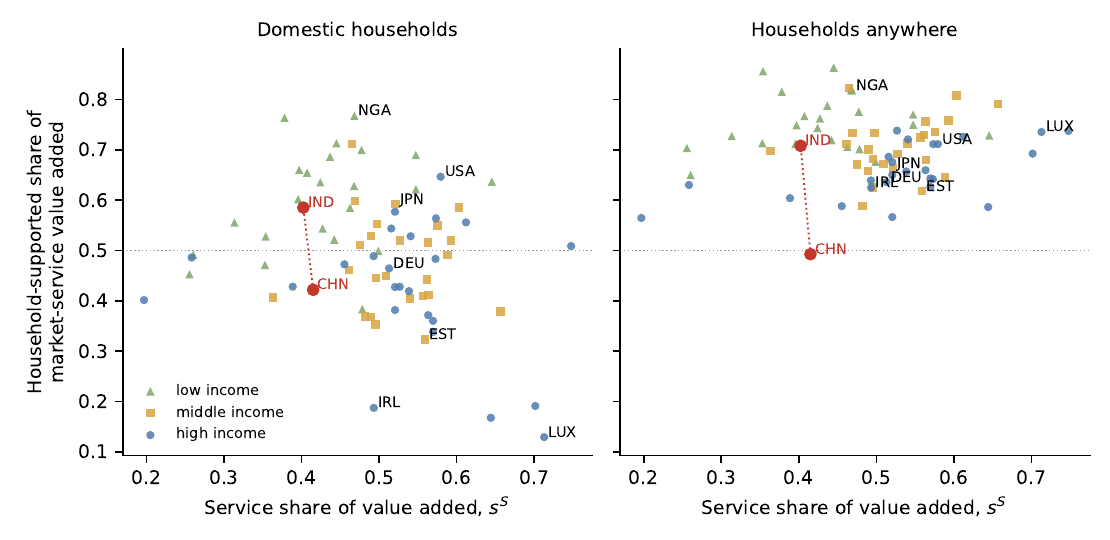}
\caption{The service share does not reveal whose demand supports
services. Market-service share of value added ($s^S$) against the
household-supported share of market-service value added, 76
economies, 2018, by income tercile. Left: domestic households only
($R$); the correlation is $-0.27$ ($t=-2.4$), and $\RNpcDom$
controlling for income; 93 country pairs differ by less than one point
of $s^S$ and more than 10 points of $R$. Right: households anywhere
($R^{\mathrm{any}}$; Section~\ref{sec:global}); 75 of the 76 economies
lie above one-half, and the partial correlation controlling for income
is $\RNpcGlob$. China and India in red.}
\label{fig:twotypes}
\end{figure}

The benchmark can be stated exactly. Across the 2018 cross-section,
\[
\operatorname{cov}(\log s^S, \log R)
= \underbrace{\operatorname{cov}(\log s^S, \log \alpha_C)}_{+\RNcovSA}
+ \underbrace{\operatorname{cov}(\log s^S, \log \sigma_C)}_{+\RNcovSSig}
- \underbrace{\operatorname{var}(\log s^S)}_{\RNvarS},
\]
so the negative raw correlation in Figure~\ref{fig:twotypes} is carried
by the mechanical denominator term, partly offset by the rising service
content of household demand through $\sigma_C$. The correlation is
$\rho = -0.27$ in levels, with a classical $t$ of $-2.4$, and $-0.25$ in
logs; the levels slope is $-\RNlevSlope$, robust s.e.\
$\RNlevSlopeSE$.

The informative term is the first one. The service share is
\textit{orthogonal} to the household share of value added:
$\operatorname{corr}(\log s^S, \log \alpha_C) = +0.01$. Yet $\alpha_C$
accounts for almost all of the cross-country variance in $\log R$ (covariance share $\RNcovA$;
Section~\ref{sec:determinants}). The statistic used to summarize
tertiarization is unrelated to the component that determines who
supports it. Conditional on income the $s^S$--$R$
relationship disappears entirely as well: the partial correlation
controlling log GDP per capita is $\RNpcDom$ ($t=0.4$). So the service
share fails as a statistic for $R$, but not because its raw correlation
is negative, which is partly arithmetic. It fails because, with or
without an income control, it shows no detectable relation to who
supports the service sector.

The failure is specific to domestic households. For $R^{\mathrm{any}}$,
which counts households anywhere, the same partial correlation is
$\RNpcGlob$ in 2018. It lies between $\RNpcGlobMin$ and $\RNpcGlobMax$
in every year from 1995 to 2018, against $\RNpcDomMin$ to $\RNpcDomMax$
on the domestic basis. Given income, a larger service share does signal
that households ultimately buy more of an economy's services. It does
not signal that the economy's own households do. The gap between the
two is the trade route, which Section~\ref{sec:global} measures.

Neither the raw nor the income-adjusted result is an artifact of the
year or of the service-sector definition. Appendix
Figure~\ref{fig:rhotime} repeats the calculation for every year from 1995 to 2022. The raw
correlation is negative in all of them (mean $-0.24$, strengthening to
$-0.34$ by 2022). The 95 percent confidence interval of the
income-adjusted partial correlation contains zero in every year.

Appendix~\ref{app:definitions} repeats the main results under four
alternative definitions of the service sector: two broader ones, one
without finance and real estate, and the narrower consumer-facing
subset. The raw correlation stays negative under the first three and
is small and insignificant for consumer-facing services. The China--India contrast and the episode result survive all
of them. Under the conventional all-services definition the household
reading fares worse, not better. The fraction of majority-household
episodes falls from 0.41 to 0.24, because education and health are
mostly government-supported. The market-service definition used here
is, if anything, the one most favorable to the consumer-upgrading
interpretation.

\begin{table}[H]
\centering
\begin{threeparttable}
\caption{Similar service shares, different domestic household shares:
selected countries, 2018}
\label{tab:canonical}
\small
\begin{tabular}{llcccc}
\toprule
$s^S$ band & Economy & $s^S$ (percentile) & $R$ & ($R$ percentile) & $R_I$ \\
\midrule
Near 0.41 & China   & 0.415 (20) & 0.421 & (28) & 0.323 \\
& India   & 0.402 (17) & 0.585 & (75) & 0.163 \\
& C\^ote d'Ivoire & 0.407 (18) & 0.654 & (89) & 0.174 \\
\midrule
Near 0.47 & Korea   & 0.455 (28) & 0.472 & (42) & 0.206 \\
& Nigeria & 0.468 (34) & 0.768 & (100) & 0.135 \\
& Thailand & 0.478 (41) & 0.383 & (18) & 0.098 \\
\midrule
Near 0.52 & Germany & 0.513 (55) & 0.460 & (38) & 0.128 \\
& Sweden & 0.520 (58) & 0.381 & (17) & 0.155 \\
& Japan   & 0.521 (62) & 0.591 & (79) & 0.182 \\
\midrule
Near 0.56 & Brazil  & 0.547 (70) & 0.689 & (93) & 0.090 \\
& Estonia & 0.559 (74) & 0.323 & (7)  & 0.111 \\
& Belgium & 0.563 (78) & 0.369 & (14) & 0.101 \\
\bottomrule
\end{tabular}
\begin{tablenotes}
\footnotesize
\item $s^S$: market-service VA over total VA; $R$: share of
market-service VA supported by domestic household consumption;
$R_I\equiv V_I^S/V^S$: investment-supported share. Parentheses give
percentiles in the 76-economy 2018 distribution. Within each band, the
largest difference in $s^S$ is 2.3 percentage points, while the spread
in $R$ ranges from 21.0 to 38.5 percentage points. At similar service
shares, investment supports nearly twice as much of China's service
value added as of India's.
\end{tablenotes}
\end{threeparttable}
\end{table}

Part of these level differences is a difference in route, and
Section~\ref{sec:global} shows how much of them survives once exports
are followed abroad. Much of the low-$R$ group exports services, or
goods that embody them, to foreign households. Counting households
anywhere narrows the interquartile range across economies from
\RNiqrLo--\RNiqrHi{} for $R$ to \RNiqrAnyLo--\RNiqrAnyHi{} for
$R^{\mathrm{any}}$.

\subsection{The fall in the domestic household share}\label{sec:trajectories}

\begin{figure}[htbp]
\centering
\includegraphics[width=0.78\textwidth]{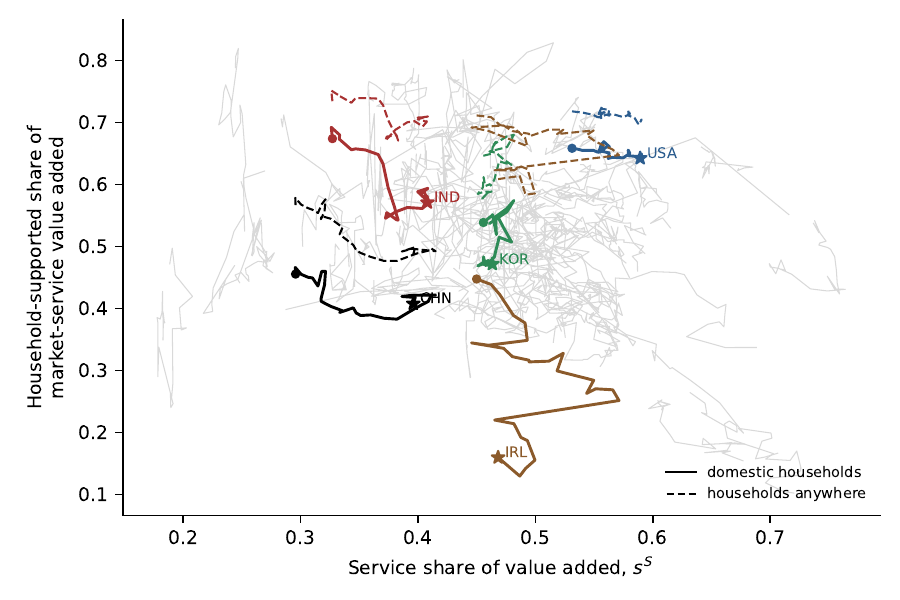}
\caption{Paths of the service share and the household-supported share
of market-service value added, 1995--2022. Solid lines count domestic
households ($R$); dashed lines count households anywhere
($R^{\mathrm{any}}$; Section~\ref{sec:global}). Dots mark 1995 and
stars 2022; gray lines
show the domestic paths of the remaining economies. Most economies
moved right and down on the domestic basis; the dashed paths lie
higher and move less.}
\label{fig:trajectories}
\end{figure}

\begin{table}[htbp]
\centering
\begin{threeparttable}
\caption{Rising service shares and the domestic household share,
1995--2018}
\label{tab:twotypes}
\small
\begin{tabular}{lrrr}
\toprule
& Raw signs & 1 pp threshold & 2 pp threshold\\
\midrule
$R$ rose
  &15 (23\%)&14 (23\%)&7 (13\%)\\
No material change in $R$
  &0 (0\%)&1 (2\%)&7 (13\%)\\
$R$ fell
  &50 (77\%)&47 (76\%)&42 (75\%)\\
\midrule
Economies with a rising service share
  &65&62&56\\
\bottomrule
\end{tabular}
\begin{tablenotes}
\footnotesize
\item Each column applies the stated cutoff to both $\Delta s^S$ and
$\Delta R$. The raw-sign column includes every economy with
$\Delta s^S>0$; the threshold columns require $\Delta s^S>.01$ or
$.02$ and classify $R$ as rising, stable within the band, or falling.
The eleven remaining economies all saw both $s^S$ and $R$ fall.
\end{tablenotes}
\end{threeparttable}
\end{table}

The rise of services after 1995 was not carried by domestic
households. Among the 65 economies whose service share rose over
1995--2018, 50 (77 percent) saw the domestic household share of their
services fall (Table~\ref{tab:twotypes} and
Figure~\ref{fig:trajectories}). Three-quarters still do when both
changes must exceed one or two percentage points, and 84 percent do in
chain volume (Section~\ref{sec:prices}). The count needs no
decomposition; it is visible once $R$ is measured.

The fall in $R$ is not confined to economies whose service share rose.
$R$ fell in \RNfellAll{} of the 76 economies, including all eleven
whose service share did not rise, where the median fall was
\RNmedDRothers{} points against 5.8 among the 65. Across economies, the
change in $R$ is unrelated to the change in the service share (slope
$-\RNslopeDRDs$, robust s.e.\ $\RNslopeDRDsSE$). So the fall is not a
by-product of tertiarization. Two forces produce it, a period effect
and development, and Table~\ref{tab:period} separates them.

The period effect has the timing of trade integration. The median $R$
fell from \RNmedRninetyfive{} in 1995 to \RNmedRoeight{} in 2008 and
\RNmedReighteen{} in 2018. So \RNshareByOeight{} percent of the fall in
the median came before 2008 (\RNvolShareByOeight{} percent in chain
volume, Panel~C of Table~\ref{tab:period}). Service shares kept rising
after 2008, in
61 of the 76 economies in each period. But among those economies the
median fall in $R$ shrank from 4.5 points in 1995--2008 to 0.8 points
in 2008--2018. The export-supported share of service value added moved
in step: it rose a median 5.6 points in the first period and 1.1 points
in the second.

The development force shows in the cross-section of changes. Panel~B of
Table~\ref{tab:period} splits each economy's change in $\log R$ into a
shift common to all economies in a period and a part proportional to
its own income growth. The common shift is $-\RNcommonA$ (s.e.\
$\RNcommonAse$) in 1995--2008 and $-\RNcommonB$ ($\RNcommonBse$) in
2008--2018, and each log point of income growth lowers $\log R$ by
\RNincSlope{} (\RNincSlopeSE). At the point estimates the two parts are
of similar size: of the average fall in $\log R$ over 1995--2018,
\RNmeanDlogR, the income part is \RNincomePart, a share of
\RNincShare. The timing and the sign are robust: most of the fall came
before 2008, and faster-growing economies saw larger falls. The split
between the two parts is less precise. The 95 percent interval of that
share runs from \RNincShareLo{} to \RNincShareHi, and in chain volume
(Panel~C) the share is \RNvolIncShare, with a common shift that
continues after 2008 ($-\RNvolCommonB$, s.e.\ $\RNvolCommonBse$).
Annual data agree. With country and year effects, the
within-country income gradient of $\log R$ is $-\RNwithinR$ (s.e.\
$\RNwithinRse$); without the seven fastest-growing Asian economies it
is similar but imprecise ($-\RNwithinRnoAsia$, s.e.\
$\RNwithinRnoAsiaSE$). Across economies in 2018 the gradient is
$-\RNbetweenR$ ($\RNbetweenRse$). It survives a control for population
($-\RNbetweenRpop$, $\RNbetweenRpopSE$), so it is not just that small
economies are both richer and more open.

\begin{table}[htbp]
\centering
\begin{threeparttable}
\caption{The fall in the domestic household share, by period}
\label{tab:period}
\footnotesize
\begin{tabular}{lccc}
\toprule
& 1995--2008 & 2008--2018 & 1995--2018 \\
\midrule
\multicolumn{4}{l}{\textit{Panel A: the fall in the domestic household share $R$}} \\
Economies with a rising service share & 61 & 61 & 65 \\
\quad $R$ fell, rising service share & 53/61 & 38/61 & 50/65 \\
\quad $R$ fell, other economies & 13/15 & 11/15 & 11/11 \\
Median $\Delta R$, rising service share (points) & $-4.5$ & $-0.8$ & $-5.8$ \\
Median change in export-supported share, all (points) & $+5.6$ & $+1.1$ & $+6.9$ \\
\midrule
\multicolumn{4}{l}{\textit{Panel B: common and income parts of $\Delta\log R$}} \\
Common shift & $-0.059$ (0.026) & $-0.007$ (0.015) & $-0.066$ \\
Income part (slope $\times$ mean $\Delta\log y$) & $-0.044$ & $-0.019$ & $-0.063$ \\
Mean $\Delta\log R$ & $-0.103$ & $-0.026$ & $-0.129$ \\
\midrule
\multicolumn{4}{l}{\textit{Panel C: chain volume, 1995 prices}} \\
Economies with a rising service share & 63 & 59 & 64 \\
\quad $R$ fell, rising service share & 56/63 & 38/59 & 54/64 \\
\quad $R$ fell, other economies & 10/13 & 11/17 & 10/12 \\
Median $\Delta R$, rising service share (points) & $-5.9$ & $-1.6$ & $-7.9$ \\
Common shift in $\Delta\log R$ & $-0.074$ (0.024) & $-0.030$ (0.014) & $-0.105$ \\
Income part & $-0.044$ & $-0.019$ & $-0.063$ \\
\bottomrule
\end{tabular}
\begin{tablenotes}
\footnotesize
\item Panel A: national tables, 76 economies. A rising service share
means $\Delta s^S>0$ over the period; the export-supported share is
$V_X^S/V^S$. Panel B: stacked long differences over 1995--2008 and
2008--2018 for the 75 economies with income data,
$\Delta\log R_{cp}=a_p+b\,\Delta\log y_{cp}$, with $y$ GDP per capita
at PPP and standard errors clustered by economy. The common shift is
$a_p$; the income part is $\hat b$ times the period's mean
$\Delta\log y$, so the two add to the mean change; the slope $\hat b$
is $-\RNincSlope$ (s.e.\ $\RNincSlopeSE$). The 1995--2018
column adds the two periods. Panel C repeats Panels A and B in chain
volume at 1995 prices (Section~\ref{sec:prices}). Its income slope is
$-\RNvolIncSlope$ (s.e.\ $\RNvolIncSlopeSE$), equal to the
current-price slope at three decimals, so the income parts coincide.
Its mean change is the sum of its two parts.
\end{tablenotes}
\end{threeparttable}
\end{table}

Section~\ref{sec:global} shows where the support went, and
Section~\ref{sec:determinants} shows which margin of the identity
produces the fall.

\subsection{Tertiarization episodes: who supports service-sector
growth}\label{sec:classification}

Episodes ask who supported the new service value added. I use three
non-overlapping windows: 1995--2005, 2005--2015, and
2015--2022. For economy $c$ and window $e=[t_0,t_1]$, the
country-window is a tertiarization episode when both
$s^S_{c,t_1}>s^S_{c,t_0}$ and $V^S_{c,t_1}>V^S_{c,t_0}$. The
76-economy panel supplies at most one observation per economy in each
window. These conditions yield 61, 63, and 37 episodes, or 161
country-window observations drawn from 73 distinct economies. Within
each episode,
$H_{ce}=\Delta V_{C,ce}^S/\Delta V_{ce}^S$ measures the household
contribution to the market-service value-added expansion. $H$ is
measured in current U.S. dollars, as in the source tables, and in
national currency, which removes exchange-rate movements between the
two endpoints. The continuous $H$ is the primary object; $H>0.5$ is
only a majority convention. Table~\ref{tab:H} reports both.

\begin{table}[H]
\centering
\begin{threeparttable}
\caption{The household contribution to tertiarization: continuous $H$
over tertiarization episodes}
\label{tab:H}
\footnotesize
\setlength{\tabcolsep}{3.5pt}
\begin{tabular}{lccccccc}
\toprule
& \multicolumn{5}{c}{Current U.S.\ dollars} & \multicolumn{2}{c}{National currency} \\
\cmidrule(lr){2-6}\cmidrule(lr){7-8}
& $n$ & Median & IQR & $H>0.5$ & Value-wtd. & Median & $H>0.5$ \\
\midrule
Pooled       & 161 & 0.453 & [0.314, 0.575] & 0.41 & 0.532 & \RNHtabPoolMed & \RNHtabPoolMaj \\
\midrule
1995--2005   & 61  & 0.478 & [0.384, 0.572] & 0.41 & 0.555 & \RNHtabWaMed & \RNHtabWaMaj \\
2005--2015   & 63  & 0.483 & [0.315, 0.591] & 0.48 & 0.514 & \RNHtabWbMed & \RNHtabWbMaj \\
2015--2022   & 37  & 0.295 & [0.193, 0.557] & 0.30 & 0.538 & \RNHtabWcMed & \RNHtabWcMaj \\
\midrule
Bottom tercile & 54  & 0.573 & [0.491, 0.643] & 0.70 & 0.483 & \RNHtabTloMed & \RNHtabTloMaj \\
Middle tercile & 53 & 0.427 & [0.290, 0.506] & 0.28 & 0.532 & \RNHtabTmidMed & \RNHtabTmidMaj \\
Top tercile  & 53  & 0.351 & [0.191, 0.498] & 0.25 & 0.554 & \RNHtabThiMed & \RNHtabThiMaj \\
\bottomrule
\end{tabular}
\begin{tablenotes}
\footnotesize
\item $H$ $=$ household contribution to the market-service VA
expansion within episodes in which both nominal market-service VA and
its share rise. Median: median $H$; $H>0.5$: fraction of episodes
with $H>0.5$; Value-wtd.: value-weighted mean $H$. The panel covers 76
economies; $n$ counts qualifying country-windows, and the pooled 161
episodes come from 73 distinct economies. Income groups: within-sample
terciles of episode-start GDP
pc (PPP), not World Bank income classes; the tercile rows sum to 160
rather than 161 because Taiwan's 1995--2005 episode has no comparable
income series and is dropped from every income cut. IQR is the interquartile
range, from the 25th to the 75th percentile. Value weights: size of
the service-VA expansion (current USD). National currency: both
endpoints converted with the World Development Indicators conversion
factor implied by services value added in current local currency and
current dollars; it covers the \RNHlcuN{} episodes with a conversion
factor at both endpoints (the windows contribute \RNHtabWaN, \RNHtabWbN,
and \RNHtabWcN; the terciles \RNHtabTloN, \RNHtabTmidN, and
\RNHtabThiN).
\end{tablenotes}
\end{threeparttable}
\end{table}

The middle half of episodes spans $H=0.314$ to $0.575$ and so crosses
the majority threshold. Income shifts this central mass. For the bottom
income tercile the interquartile range is $[0.491,0.643]$; for the top
tercile it is $[0.191,0.498]$. Regressing $H$ on log income at the
episode start, with window fixed effects and standard errors clustered
by country, gives $-0.098$ (s.e.\ $0.021$; 160 episodes in 72
economies). In national currency it gives $-\RNHgradLcu$ (s.e.\
$\RNHgradLcuSE$; \RNHlcuN{} episodes). Controlling for each economy's
1995--2018 relative service-price change leaves the gradient unchanged
($-0.113$ against $-0.115$ in the same 142-episode subsample). The
typical low-income episode is majority household (median $H=0.57$),
the typical high-income one is not (median 0.35). The gradient mostly
restates the level gradient of $R$. Since $H=(gR_1-R_0)/(g-1)$, with
$g$ the episode's growth factor of service value added, $H$ is close
to the end-of-episode $R$. Controlling for that $R$, the gradient falls
to $-\RNHgradRone$ (s.e.\ $\RNHgradRoneSE$). Large high-income episodes
are more household-supported: the three United States episodes have
$H$ of 0.60--0.66, so the gradient is weaker under value weighting.

Four episodes have $H$ outside $[0,1]$ in dollars: Hungary, Iceland,
and the Netherlands in 2005--2015 and Myanmar in 2015--2022. All four
lie inside $[0,1]$ in national currency, so they are exchange-rate
artifacts. Dropping them gives a median $H$ of 0.457 and 42 percent
majority household. In national currency the median is \RNHlcuMed{}
and \RNHlcuMaj{} percent of episodes are majority household, against
\RNHusdMed{} and \RNHusdMaj{} percent in dollars on the same episodes;
\RNHflips{} of the \RNHlcuN{} majority labels change. The 2015--2022
window overlaps the pandemic; split at 2019, the windows 2015--2019
and 2019--2022, each with its own episodes, both give a median $H$ of
0.36. Other window definitions give a median $H$ of
0.44 to 0.48 (Appendix Table~\ref{tab:windows}). Among the 95 episodes with $H<0.5$, exports are
the largest non-household contributor in 91 and investment in four.
That is the first sign of what Section~\ref{sec:global} shows: on the
domestic basis, the main competitor of domestic households in service
growth is foreign demand.

\section{Following Exports to Their Final Users}\label{sec:global}

The domestic tracing treats exports as final. But an exported service,
or a good that embodies one, is finally absorbed by someone: a foreign
household, a foreign government, or a foreign investor. The OECD
inter-country tables let me follow it. I trace each economy's
market-service value added through the full inter-country Leontief
inverse, 81 regions by 50 industries, to final demand anywhere in the
world. I record it by destination economy and by type of final user.
Households are household consumption wherever it occurs, including
purchases by non-residents. $R^{\mathrm{any}}$ is the share of an
economy's market-service value added that these households support.
The traced value added equals each
economy's own market-service value added to within $10^{-14}$ in every
year. Restricted to final users at home, the measure reproduces the
domestic $R$ (correlation 0.99998; the small gap is value added that
is exported and later re-imported).

\begin{table}[htbp]
\centering
\begin{threeparttable}
\caption{Domestic households and households anywhere}
\label{tab:global}
\small
\begin{tabular}{lcccc}
\toprule
 & \multicolumn{2}{c}{Current prices} & \multicolumn{2}{c}{Chain volume} \\
\cmidrule(lr){2-3}\cmidrule(lr){4-5}
Households counted & Domestic & Anywhere & Domestic & Anywhere \\
\midrule
Median $R$, 2018 & 0.50 & 0.71 & -- & -- \\
Economies with $R>0.5$, 2018 & 37 & 75 & -- & -- \\
Rising $s^S$, falling $R$, 1995--2018 & 50/65 (77\%) & 47/65 (72\%) & 54/64 (84\%) & 42/64 (66\%) \\
\quad both changes above 2 points & 41/56 (73\%) & 30/56 (54\%) & 40/59 (68\%) & 31/59 (53\%) \\
Median $\Delta R$ if $s^S$ rose (points) & $-$5.8 & $-$2.6 & $-$7.9 & $-$2.3 \\
Median $H$, 161 episodes & 0.453 & 0.678 & 0.415 & 0.667 \\
Share of episodes with $H>0.5$ & 42\% & 96\% & 36\% & 93\% \\
\bottomrule
\end{tabular}
\begin{tablenotes}
\footnotesize
\item ``Domestic'' counts only the economy's own households;
``anywhere'' adds households abroad, including non-residents'
purchases in the economy. Changes and counts use economies whose
market-service share rose over 1995--2018. $H$ holds the 161
current-price episodes fixed. Chain volumes are chain-linked from the
previous-year-price tables (Section~\ref{sec:prices}). The domestic
columns differ slightly from Tables~\ref{tab:twotypes}, \ref{tab:H},
and~\ref{tab:prices} because this tracing includes re-imported value
added. \textit{Source}: OECD inter-country
input-output tables in current and previous-year prices, 2026 release;
76 economies.
\end{tablenotes}
\end{threeparttable}
\end{table}

Foreign households absorb most of what the domestic tracing calls
export support. At the median, foreign final demand supported 23
percent of an economy's market-service value added in 1995 and 29
percent in 2018. Of that, foreign households absorb 65 percent in 2018
(69 percent in 1995), foreign investment 27 percent, and foreign
government 7 percent. The destinations are spread out. At the median,
the European Union takes 28 percent, the United States 11 percent,
China 8 percent, and all other economies 48 percent; medians need not
add up.

Counting households anywhere changes the levels sharply
(Table~\ref{tab:global}). In 2018 the median share of market-service
value added that households support rises from 0.50 for $R$ to 0.71 for
$R^{\mathrm{any}}$. Households anywhere support a majority of services
in 75 of the 76 economies, against 37 on the domestic basis. The
exception is China, at 0.49, because domestic and foreign investment
together support 37 percent of its services; the figure for India is 24
percent. On this basis, 96 percent of tertiarization episodes are
majority household.

The change survives in direction but shrinks. Among economies whose
service share rose, $R^{\mathrm{any}}$ fell in 47 of 65 (72 percent),
against 50 on the domestic basis. But only 30 of 56 meet the two-point
threshold, and the median fall is 2.6 points instead of 5.8.
Table~\ref{tab:destinations} shows where the domestic household share
went. Over 1995--2008 it fell a median 4.5 points; foreign households
gained 2.2 points and foreign investors, reached through exports, 2.1.
So the support that left domestic households went abroad, about
equally to foreign households and to foreign investment. $R^{\mathrm{any}}$ also fell in
that period, in \RNanyFellA{} of the 61 economies. Over 1995--2018, the
gains of foreign households and foreign investors together roughly
match the domestic households' loss (median ratio \RNfrnRatioC). Over
2008--2018 domestic households lost only 0.8 points, and
$R^{\mathrm{any}}$ gained 0.8: it fell in only \RNhhAnyLateFell{} of
the \RNrisersLate{} economies whose service share rose. The investment
that gained over the whole period was foreign. The share of services
supported by domestic investment changed by a median $\RNinvDom$ points
and rose in only \RNinvDomRose{} percent of these economies. The share
supported by investment abroad changed by $\RNinvFrn$ points and rose
in \RNinvFrnRose{} percent.

\begin{table}[htbp]
\centering
\begin{threeparttable}
\caption{Where the support for services went, by period}
\label{tab:destinations}
\small
\begin{tabular}{lccc}
\toprule
& 1995--2008 & 2008--2018 & 1995--2018 \\
\midrule
Domestic households & $-4.5$ & $-0.8$ & $-5.8$ \\
Foreign households & $+2.2$ & $+0.5$ & $+3.4$ \\
Domestic investment & $+0.4$ & $-1.1$ & $-0.9$ \\
Foreign investment & $+2.1$ & $+0.3$ & $+2.2$ \\
Government, home and abroad & $+0.1$ & $+0.2$ & $+0.1$ \\
Households anywhere & $-2.7$ & $+0.8$ & $-2.6$ \\
\midrule
Economies with a rising service share & 61 & 61 & 65 \\
\bottomrule
\end{tabular}
\begin{tablenotes}
\footnotesize
\item Median change, in percentage points, in the share of
market-service value added absorbed by each type of final user, among
economies whose service share rose over the period (national tables).
Foreign households and foreign investment are reached through exports.
Medians need not add up. \textit{Source}: OECD inter-country
input-output tables at current prices, 2026 release; 76 economies.
\end{tablenotes}
\end{threeparttable}
\end{table}

Income growth works differently. Within countries, with country and
year effects, $R^{\mathrm{any}}$ falls with income as much as the
domestic household share does: $-\RNgradAnyW$ (s.e.\ $\RNgradAnyWse$) per log
point of GDP per capita, against $-\RNgradDomW$ ($\RNgradDomWse$), both
in levels (in logs, $-0.083$ and $-0.088$). Growth moves support from
households to investment. With the same design, the share of services
supported by investment at home or abroad rises \RNgradInvAnyW{} (s.e.\
\RNgradInvAnyWse) per log point, and the share supported by domestic
investment alone rises \RNgradInvDomW{} (\RNgradInvDomWse). The
investment share of value added rises \RNdetWshIabs{} per log point
(Section~\ref{sec:determinants}). Across
economies, the
level gradient of $R^{\mathrm{any}}$ in 2018 is smaller,
$-\RNgradAnyB$ ($\RNgradAnyBse$), than that of $R$, $-\RNgradDomB$
($\RNgradDomBse$): richer
economies sell more of their services abroad, mostly to households.

The two bases answer different questions, and both answers matter.
Households remain the destination of most service output, at home or
abroad. Exports changed the route by which services reach
them, and within countries growth shifted some support from households
to investment. For the domestic economy, both are changes in whose
demand its service sector depends on. Section~\ref{sec:inference} shows
that this changes how service output moves when that demand moves.

\section{An Exact Decomposition of \texorpdfstring{$R$}{R}}\label{sec:determinants}
\label{sec:decomp}

Differences in $R$ across economies are differences in the household
share of value added, not in what households buy. Taking logs of
equation~\ref{eq:Ridentity} gives the additive form
\[
\log R_{ct} \;=\; \log \alpha_{C,ct} + \log \sigma_{C,ct} - \log s^S_{ct},
\]
which puts the three margins in common units. In the 2018
cross-section, the covariance decomposition of $\log R$ assigns
almost all of its variance to $\log \alpha_C$ (covariance share
$\RNcovA$). The other two margins, $\sigma_C$ and $-s^S$, contribute
$-\RNcovSig$ and $+\RNcovS$ and largely offset each other. So I
decompose $\alpha_C$ into the other final-demand categories.

\begin{table}[htbp]
\centering
\begin{threeparttable}
\caption{Income gradients of the components of $R$ and of the
household share of value added}
\label{tab:determinants}
\small
\begin{tabular}{lcc}
\toprule
\multicolumn{3}{l}{\textit{Panel A: log components of $R=\alpha_C\sigma_C/s^S$}} \\
\midrule
Component & Within & Between (2018) \\
\midrule
Household share of value added, $\log \alpha_C$ & $-0.1635$ (0.0322) & $-0.2653$ (0.0420) \\
Service content of household demand, $\log \sigma_C$ & $+0.1113$ (0.0340) & $+0.1689$ (0.0211) \\
Service share, $\log s^S$ & $+0.0356$ (0.0276) & $+0.1245$ (0.0327) \\
Domestic household share of services, $\log R$ & $-0.0879$ (0.0336) & $-0.2209$ (0.0451) \\
\bottomrule
\end{tabular}

\medskip

\begin{tabular}{lcc}
\toprule
\multicolumn{3}{l}{\textit{Panel B: components of $\alpha_C+\alpha_G+\alpha_I+\alpha_X=1$}} \\
\midrule
Component & Within & Between (2018) \\
\midrule
Household share of value added, $\alpha_C$ & $-0.0797$ (0.0148) & $-0.0931$ (0.0136) \\
Government share of value added, $\alpha_G$ & $-0.0084$ (0.0091) & $+0.0291$ (0.0072) \\
Investment share of value added, $\alpha_I$ & $+0.0739$ (0.0110) & $-0.0133$ (0.0055) \\
Export share of value added, $\alpha_X$ & $+0.0142$ (0.0140) & $+0.0773$ (0.0167) \\
\midrule
Sum & $0.0000$ & $0.0000$ \\
\bottomrule
\end{tabular}
\begin{tablenotes}
\footnotesize
\item Each entry is the coefficient on log GDP per capita (PPP).
Within: annual panel 1995--2018 ($n=\RNdetN$, \RNdetG{} economies),
country and year fixed effects, country-clustered SEs. Between: 2018
cross-section ($n=\RNdetBN$), robust SEs. Panel A uses logs; since
$\log R=\log\alpha_C+\log\sigma_C-\log s^S$, the first two rows minus
the third equal the fourth in the unrounded estimates. Panel B uses
share levels and the same sample and design for all four categories,
so the coefficients add exactly to zero. World medians, 2018:
$\alpha_C=0.38$, $\sigma_C=0.67$, $s^S=0.50$, $R=0.50$.
\end{tablenotes}
\end{threeparttable}
\end{table}

The decomposition separates three margins. The first is an Engel
force: development raises the service content of household demand
($\log \sigma_C$: $t=8.0$ between countries, $t=3.3$ within). The positive sign is the
demand-side prediction of nonhomothetic structural-transformation
\mbox{models}: because services have higher income elasticity, rising
income shifts household demand toward services
\citep{kongsamut2001beyond, comin2021structural,
boppart2014structural, matsuyama2019engel, herrendorf2014growth}. The
mapping is not unique, however. Because $\sigma_C$ is a current-price
production-network statistic, the same sign can also reflect
relative-price and productivity forces \citep{baumol1967unbalanced,
ngai2007structural} or changing input-output linkages. Whatever the
mechanism, the descriptive fact is clear: $\sigma_C$ rises with
income.

The other two margins, and $R$ itself, have a direct accounting
reading. Within
countries, a 1 percent increase in GDP per capita is associated with a
\RNdetWAabs{} percent decline in $\alpha_C$: household consumption
supports a smaller share of economy-wide value added. It is associated
with a \RNdetWSigabs{} percent rise in $\sigma_C$, the service content
of household demand. The aggregate service-share gradient,
$\RNdetWSabs$ (s.e.\ $\RNdetWSSE$), is not statistically
distinguishable from zero once country and year effects are removed.
Because $\log R=\log\alpha_C+\log\sigma_C-\log s^S$, these components
reconcile exactly in the unrounded estimates; in rounded numbers,
$-\RNdetWAabs+\RNdetWSigabs-\RNdetWSabs\approx-\RNdetWRabs$. So the
negative within-country gradient in $R$ means that the service sector
becomes less household-supported as income rises. The main reason is
that the fall in $\alpha_C$ outweighs the rise in $\sigma_C$.

Panel B shows where the falling household share of value added goes.
Within countries, the $-\RNdetWshCabs$ gradient in $\alpha_C$ is
matched mainly by a $+\RNdetWshIabs$ gradient in the investment share
of value added, $\alpha_I$. This is the extensive margin of structural
change in \citet{garciasantana2021investment}, the reallocation of
final expenditure between consumption and investment, measured here,
as there, in value added. It is identified from differences in growth across
economies, where fast growers had investment booms. It is not a common
trend in the support for services. Over 1995--2018 the share of
services supported by domestic investment fell in the median economy
whose service share rose (Section~\ref{sec:global}). The export
gradient is positive but imprecise, and the government share barely
moves. Across countries, the pattern is different: the
$-\RNdetBshCabs$ household gradient is matched by the export share
($+\RNdetBshXabs$) and the government share ($+\RNdetBshGabs$), while
the investment share declines modestly ($-\RNdetBshIabs$). The four
coefficients add to zero by construction. These describe the category
shares of value added. They do not say why saving, exports, or
government demand differ. Section~\ref{sec:global} shows that most
of the export-supported part ends with foreign households.

The 2018 cross-section is also stronger on the third margin. Richer
economies have lower $\alpha_C$ ($-\RNdetBAabs$), higher $\sigma_C$
($+\RNdetBSigabs$), and larger service sectors ($+\RNdetBSabs$), all
statistically significant. Since $s^S$ enters the identity negatively,
the larger aggregate service sector reinforces the decline in
$\alpha_C$, yielding the more negative between-country gradient in $R$
($-\RNdetBRabs$). In both comparisons, $\sigma_C$ rises with income,
yet the service economy as a whole tilts toward non-household demand.
The familiar Engel force is visible in $\sigma_C$, but the household
share of value added falls, and the second force outweighs the first.
The aggregate service share mixes the two; Section~\ref{sec:engel}
shows what that does to a standard demand estimate. The distinction
also survives when service consumption is scaled by household
disposable income (Appendix~\ref{app:construction}). Household budgets
tilt toward services as income rises. But the budget measure correlates
only 0.11 with $R$ across economies in 2018, against 0.67 for
$\sigma_C$.

The component levels also clarify the country comparisons, including
China and India (Appendix Table~\ref{tab:chnind}). The two have nearly identical
service shares \textit{and} a similarly modest service content of
household demand ($\sigma_C$ of 0.54 and 0.45, both below the 0.67
world median). The 16-point gap in $R$ comes almost entirely from the
household share of value added. India's household demand occupies a far
larger position in its economy ($\alpha_C = 0.52$ against China's 0.32,
the latter among the lowest in the world). So the same-sized service
sector is much more household-supported in India. The China--India
difference is not in the service content of household demand; it is
that China's aggregate demand is unusually tilted away from households. Ireland's extreme $R$, by
contrast, is entirely an $\alpha_C$ story (0.15).

The comparison is a within-year one, so prices do not affect $R$; the
components $\alpha_C$, $\sigma_C$, and $s^S$ are current-price
descriptions. Prices matter for China's \textit{changes}:
Section~\ref{sec:prices} shows that whether China's real service share
rose or fell depends on the deflator, while its $R$ falls on every
price basis.

A one-type model cannot produce these patterns. If every unit of value
added is supported by domestic household demand, then $\alpha_C=1$ and
$R=1$ in every year; nonhomothetic demand raises $\sigma_C=s^S$, but
nothing can move $R$. A second purchaser is enough to produce the
within-country pattern. Suppose households save a share of income that
rises with income, and saving finances investment. Then
investment-supported value added grows faster than
household-supported value added, $\alpha_C$ falls, and the identity
gives $R$ falling whenever $\alpha_C$ falls faster than $\sigma_C/s^S$
rises. Panel B of Table~\ref{tab:determinants} shows this
within-country margin in the data. Between countries, and in the
common shift of 1995--2008, the same role is played by exports, which
require no model of saving.

\section{Accounting for the Changes}\label{sec:accounting}

This section allocates the observed changes in $s^S$ and $R$, exactly,
to six measured components of the input--output tables. The allocation
shows which components moved and what those movements imply for the two
statistics. It does not claim that the components are exogenous or that
their Shapley contributions are causal effects.

\subsection{Accounting components and closure}

An economy with $n=50$ industries and market-service set $\mathcal S$
is represented in year $t$ by
\[
\theta_t=\bigl(\underbrace{A_t,v_t}_{\text{production block}},
b_t^C,b_t^G,b_t^I,b_t^X,\varphi_t\bigr),
\]
with six component labels
\[
\mathcal P=\bigl\{(A,v),b^C,b^G,b^I,b^X,\varphi\bigr\}.
\]
Here $A_t$ is the domestic input-coefficient matrix and $v_t$ the
vector of industry value-added coefficients. Each category's final
demand has a scale and a shape. The scale $\varphi_{k,t}$, its share of
total final demand, is one number; the shape $b_t^k$, its industry
composition, is a vector that sums to one. The four scales, $\varphi_t$, are
the \textit{final-demand structure}. The columns of the production
block obey
\[
v_{jt}+\sum_i A_{ijt}+m_{jt}=1,
\]
where $m_{jt}\geq0$ collects imported inputs and net taxes.

The accounting restriction matters for decomposition. Treating $A$ and
$v$ as separate accounting components creates infeasible synthetic
economies: 90 percent of mixed-year $(A,v)$ combinations generate a
negative residual in at least one industry column. So I treat $(A,v)$
as one production component, and every coalition evaluated below uses
a same-year production block. Appendix~\ref{app:accounting} reports the
feasibility audit of the measured blocks.

Because both outcomes are shares, normalize total final demand to one,
so $f^k=b^k\varphi_k$: shape times scale. For any $\theta$, gross
output supported by category $k$ is
\[
x^k(\theta)=(I_n-A)^{-1}b^k\varphi_k,
\]
and the two outcomes are
\[
s^S(\theta)=
\frac{\sum_{i\in\mathcal S}v_i\sum_kx_i^k}
{\sum_i v_i\sum_kx_i^k},
\qquad
R(\theta)=
\frac{\sum_{i\in\mathcal S}v_i x_i^C}
{\sum_{i\in\mathcal S}v_i\sum_kx_i^k}.
\]

The formulas reproduce the measured statistics exactly. At measured
values of $\theta_t$ they return the directly constructed $s^S_{ct}$
and $R_{ct}$ to machine precision at all 152 country--year endpoints,
without resetting Romania's small negative residual
(Appendix~\ref{app:accounting}). The mapping also
shows what a
final-consumption model disciplines: the household-demand composition
$b^C$, but not $R$, which depends on every other block as well.

\subsection{Exact channel decomposition}

The outcomes $s^S$ and $R$ are nonlinear in the six accounting
components. So a sequential decomposition, which switches one
component at a time, depends on the order of switching. I use the exact
Shapley decomposition: the unique order-free attribution that treats
the components symmetrically and sums exactly to the observed change
\citep{shapley1953value, shorrocks2013decomposition}.
Appendix~\ref{app:accounting} discusses its precedents in
distributional analysis and further methodological points.

For each economy, the 1995--2018 change in
$y\in\{s^S,R\}$ is decomposed across the six grouped accounting components using
exact Shapley values:
\[
\Delta y_c=\sum_{p\in\mathcal P}\psi_{p,c},\qquad
\psi_{p,c}=\sum_{\mathcal Q\subseteq\mathcal P\setminus\{p\}}
\frac{|\mathcal Q|!(6-|\mathcal Q|-1)!}{6!}
\bigl[y_c(\mathcal Q\cup\{p\})-y_c(\mathcal Q)\bigr].
\]
Here $y_c(\mathcal Q)$ evaluates the components in coalition $\mathcal Q$
at their 2018 values
and all remaining components at their 1995 values. The weight on each
coalition is the share of the $6!$ orderings of the components in
which exactly the members of $\mathcal Q$ precede $p$. So $\psi_{p,c}$
is $p$'s marginal contribution averaged over every order. With six
components,
exact enumeration requires $2^6$ evaluations per economy and outcome;
no sampling or approximation is involved.

\begin{table}[H]
\centering
\begin{threeparttable}
\caption{Grouped Shapley decomposition of changes in the service share
and the domestic household share, 1995--2018}
\label{tab:shapley}
\scriptsize
\setlength{\tabcolsep}{3pt}
\begin{tabular}{lrrrrrrrr}
\toprule
&\multicolumn{4}{c}{$\Delta s^S$ (median total $+.040$)}&\multicolumn{4}{c}{$\Delta R$ (median total $-.067$)}\\
\cmidrule(lr){2-5}\cmidrule(lr){6-9}
Channel&Median&IQR&Share&Sign&Median&IQR&Share&Sign\\
\midrule
Production block $(A,v)$&$.022$&[$.006,.036$]&$.47$&$.83$&$-.000$&[$-.008,.011$]&$-.02$&$.47$\\
Household-consumption composition $b^C$&$.013$&[$.004,.022$]&$.26$&$.80$&$.014$&[$.008,.026$]&$-.13$&$.33$\\
Government composition $b^G$&$-.000$&[$-.003,.002$]&$-.00$&$.47$&$.001$&[$-.001,.003$]&$-.00$&$.45$\\
Investment composition $b^I$&$.003$&[$-.001,.008$]&$.05$&$.58$&$-.003$&[$-.009,.001$]&$.03$&$.72$\\
Export composition $b^X$&$.006$&[$-.004,.023$]&$.20$&$.66$&$-.007$&[$-.025,.006$]&$.11$&$.62$\\
Final-demand structure $\varphi$&$-.004$&[$-.011,.008$]&$-.01$&$.50$&$-.068$&[$-.121,-.020$]&$.97$&$.92$\\
\midrule
Combined production system&$.032$&[$.012,.059$]&$.73$&$.91$&$-.009$&[$-.024,.012$]&$.14$&$.64$\\
\bottomrule
\end{tabular}
\begin{tablenotes}\footnotesize
\item Exact Shapley contributions across 76 economies. ``Share'' is
the median of $\psi_{p,c}/\Delta y_c$ for economies with
$|\Delta y_c|>.02$ ($n=59$ for $s^S$, $68$ for $R$); ``Sign'' is the
fraction whose channel contribution has the sign of the economy's total
change. Medians of components need not sum to the median total;
components sum exactly within every economy. The ``Combined production
system'' row sums $\psi^{(A,v)} + \psi^{b^I} + \psi^{b^X}$
within each economy \textit{before} computing medians, shares, and
sign agreement; it is the sum of the three channels' Shapley values,
not the Shapley value of a merged channel. The group contains the
production block and the
\textit{composition} of investment and export demand; the aggregate
shares of those categories remain in $\varphi$, so the group is not
the complete production-side contribution in an economic sense.
Among economies whose service share
rose materially ($\Delta s^S > .02$, $n=56$) the median country-level
combined share is $.73$ (value-weighted $.73$, positive in 98
percent); among economies with material $R$ declines
($\Delta R < -.02$, $n=57$) the median $\varphi$ share is $1.00$
(value-weighted $1.07$, negative contribution in all 57). Further
robustness (cutoff sensitivity, percentile ranges, alternative
groupings) is in Appendix~\ref{app:accounting}. These are accounting
allocations of observed changes across measured components, not
estimates of causal mechanisms.
\end{tablenotes}
\end{threeparttable}
\end{table}

The decomposition gives two accounting results (Table~\ref{tab:shapley}).
First, most of the growth in the service share is allocated to the
production system, which groups the production block with the industry
composition of investment and export demand. The group does not
identify a production-side mechanism, and its size depends on the
grouping. The combined channel $\psi^{\mathrm{prod}}_c =
\psi^{(A,v)}_c + \psi^{b^I}_c + \psi^{b^X}_c$ sums each economy's
contributions \textit{before} any ratio is taken. Among the 56
economies whose shares rose materially, it receives a median
country-level share of 73 percent of the service-share rise. The
interquartile range is 56 to 96 percent, and the value-weighted
aggregate is also 73 percent. It has the sign of the total in 91
percent of all economies. Among the same 56 economies the production
block alone receives a median \RNpbShareFiftysix{} percent, household
composition 26 percent, and the four final-demand compositions
together \RNallCompFiftysix{} percent. So a split into production
coefficients and the composition of final demand gives roughly half to
each.

Second, the decline in $R$ is allocated almost entirely to the
final-demand structure $\varphi$. Among the 57 economies with material
$R$ declines, the median country-level $\varphi$ share is 1.00, and the
$\varphi$ contribution is negative in every one. Changes in
household-consumption composition, $b^C$, typically raise $R$ but are
overwhelmed by the falling household share of final demand. This second
result is the identity's $\alpha_C$ margin in another form: across
economies, the $\varphi$ contribution to $\Delta R$ correlates 0.96
with the change in $\alpha_C$.

Inside the production block, most of the allocation goes to the input
mix (Appendix~\ref{app:productionblock}). Among the 56 economies,
greater use of market-service inputs receives a median 35 percent of
the service-share rise. Business services are the largest single
input, at 16 percent, and lower use of goods inputs adds 11 percent.
Value-added coefficients and the domestic share of inputs each net out
near zero. In chain volume, information and communication replace
business services as the leading input.

The weight of production linkages in the first result has precedents.
\citet{sposi2019evolving} finds that cross-country differences in
input-output linkages account for 74 percent of the curvature of
industry's hump-shaped share of value added. That is twice as much as
the composition of final demand accounts for. \citet{foerster2025past}
attribute the United States' shift from goods to services mainly to two
changes. Intellectual property products replaced durable goods as
inputs, and consumer spending shifted from nondurable goods to
services.

The country-level allocations contain a second pattern. Among the 56
economies with material service-share increases, the production system
is the largest of four exhaustive grouped contributions in 45
(80 percent). So the median result is not driven by a small number of
large allocations. Its importance also rises across the development
distribution (Appendix Table~\ref{tab:mechanismmoments}). The median
production-system share increases from .57 in the low-income tercile
to .91 in the high-income tercile, while the household-composition
share falls from .50 to .13. These are cross-country accounting
moments, not estimates of an income effect: income, production
structure, and demand composition are jointly determined.

The production block's contribution to $\Delta R$ has the same sign as
the total change in only 47 percent of economies, against 83 percent
for the service share. Appendix~\ref{app:composition} shows why. To
first order, greater use of service inputs raises $R$ only in one case.
The industries that use more of them must sell more of their output to
domestic households, directly and through supply chains, than the
service sector as a whole does. That condition signs the response of
$R$ to the
observed 1995--2018 change in service input use correctly in all 76
economies.

\section{Results in Chain Volume}\label{sec:prices}

Most results so far use current-price value added, so changes mix real
reallocation with relative-price movement. The OECD's inter-country
input-output tables in previous-year prices (2026 release) cover the
same economies and the same 50 industries, so every headline statistic
can be recomputed in chain volume. The current-price version of those
tables reproduces the national-table measures used above
(Section~\ref{sec:framework}). So the differences in
Table~\ref{tab:prices} come from prices alone.

\begin{table}[htbp]
\centering
\begin{threeparttable}
\caption{Headline results at current prices and in chain volume}
\label{tab:prices}
\small
\begin{tabular}{lccc}
\toprule
 & Current & \multicolumn{2}{c}{Chain volume} \\
\cmidrule(lr){3-4}
 & prices & 1995 prices & 2018 prices \\
\midrule
Rising $s^S$ and falling $R$, 1995--2018 & 50/65 (77\%) & 54/64 (84\%) & 54/64 (84\%) \\
\quad both changes above 2 points & 42/56 (75\%) & 40/59 (68\%) & 41/59 (69\%) \\
Signs that differ from current prices: $\Delta s^S$; $\Delta R$ & -- & 15; 9 & \\
Median $H$, paper's 161 episodes & 0.453 & 0.415 & \\
Share of episodes with $H>0.5$ & 41\% & 36\% & \\
Production-system share of $\Delta s^S$ (median) & 73\% & 67\% & \\
\quad production block alone & 48\% & 42\% & \\
\quad largest group, share of economies & 80\% & 76\% & \\
Final-demand-structure share of $\Delta R$ (median) & 97\% & 90\% & \\
\midrule
China: $\Delta s^S$; $\Delta R$ (points) & +11.9; $-$3.4 & $-$6.2; $-$7.7 & $-$11.0; $-$8.5 \\
India: $\Delta s^S$; $\Delta R$ (points) & +7.5; $-$9.0 & +19.2; $-$20.0 & +14.9; $-$24.7 \\
\bottomrule
\end{tabular}
\begin{tablenotes}
\footnotesize
\item Chain volumes link each year's previous-year-price table to the
preceding current-price table. Each aggregate is chained separately,
from a 1995 reference; the last column re-references the same chains
to 2018 prices. $H$ holds the 161 current-price episodes fixed. The
current-price Shapley column is the single 1995--2018 transition of
Table~\ref{tab:shapley}; the volume column sums the adjacent-year
transitions over 1995--2018 (the adjacent-year current-price sum gives
74 percent, largest group in 79 percent, and 97 percent). The
production block's 48 percent is the median among the 56 economies
whose service share rose materially; Table~\ref{tab:shapley}'s .47
covers the 59 with a material change in either direction.
\textit{Source}: OECD inter-country input-output tables in current
and previous-year prices, 2026 release; 76 economies.
\end{tablenotes}
\end{threeparttable}
\end{table}

The fall in $R$ among economies with a rising service share survives
price adjustment and is stronger on raw signs. In chain volume, $R$
fell in 54 of the 64 economies whose service share rose, 84 percent, at
either reference year. Requiring both changes to exceed two points
lowers the fraction to 68--69 percent, against 75 at current prices.
Households contribute less to service growth in volume. The median $H$
across the 161 episodes falls from 0.453 to 0.415, and the share of
majority-household episodes from 41 to 36 percent. The Shapley
allocation keeps its ordering. The production system receives a median
67 percent of service-share growth in volume, against 73 percent at
current prices, and remains the largest group in three quarters of
economies. The final-demand structure receives 90 percent of the change
in $R$, against 97.

The domestic household share is more robust to prices than the
service share.
Price adjustment changes the sign of the 1995--2018 change in $s^S$ in
15 economies and the sign of the change in $R$ in 9
(Appendix Table~\ref{tab:signchanges} lists them). Across economies, the
current-price and volume changes correlate \RNcorrDRCpVol{} for $R$
but only \RNcorrDsCpVol{} for $s^S$. Levels are not affected at all. In
any year, $R$ in volume at that year's prices equals $R$ at current
prices, so the cross-sectional results of Section~\ref{sec:reveal} do
not depend on the price basis.

The volume measures inherit the construction of the
previous-year-price tables. The OECD deflates each industry's gross
output and value added and the total of each final-demand category. It
then estimates the remaining cells, including final demand by
supplying industry, with a RAS procedure \citep{guilhoto2022pyp}. Each
cell ends up with its own implicit deflator, but prices that differ
across uses \citep{gaggl2026structural} enter only through those
industry and final-demand totals. Because the tables are converted at
the previous year's exchange rate, the volume $H$ is measured at
constant prices and constant exchange rates. For a few economies the
volume changes are large. India's $R$ falls 9 points at current prices
but 20 to 25 points in volume. The volume and current-price changes in
$R$ differ by more than 15 points for Nigeria, Belarus, and Cambodia.

Because $R$ is a within-year ratio, I treat its levels and
current-price changes as the baseline and volume changes as a
robustness check. A deflator audit supports this choice
(Appendix~\ref{app:deflators}). The volume counts do not depend on the
eight economies whose volume paths disagree with national accounts.
Most of the volume gap in $R$ comes from prices that differ across
uses, which national accounts do not publish.

China is where prices, and the choice of deflator, matter most. Its
market-service share rises 11.9 points at current prices but falls 6.2
points in chain volume at 1995 prices and 11.0 points at 2018 prices.
In the inter-country tables the price of services relative to industry
rises about fivefold, against 81 percent in the World Development
Indicators. The gap comes mainly from how manufacturing value added is
deflated. With single deflation, China's market-service share rises 1.4
points in volume instead of falling 6.2 (Appendix~\ref{app:deflators}).
So whether China's real service share rose or fell depends on how
industry value added is deflated, and the paper does not use China as
evidence of real servicification. Both sources agree that relative
prices account for most or all of the nominal rise. China's $R$, in
contrast, falls on every basis: by 3.4 points at current prices, 7.7
at 1995 prices, and 8.5 at 2018 prices.

The China level comparisons in Table~\ref{tab:canonical} and
Appendix Table~\ref{tab:chnind} are unaffected, since they are within-year
ratios. The separate components $\alpha_C$, $\sigma_C$, and $s^S$ of
those tables are still current-price descriptions. Their denominators
mix service and non-service value added, so a relative-price change
moves each of them even when $R$ is unchanged.

\section{Why Whose Demand Matters}\label{sec:inference}

The domestic household share measures whose demand a country's
services depend on. This section shows two consequences. First,
services inherit the volatility of the demand that supports them, so
whose demand supports them changes how they move with the cycle
(Section~\ref{sec:exposure}). Second, a model that reads the service
share as household demand finds a weaker income effect than the
service content of household demand shows (Section~\ref{sec:engel}).

\subsection{Whose demand moves services}\label{sec:exposure}

On the domestic basis, foreign demand supported a median \RNexpShare{}
percent of the growth in market-service value added during
tertiarization episodes, in dollars. The median is \RNexpLcuMed{}
percent in national currency and \RNexpVolShare{} percent in chain
volume. Foreign demand supported a majority of that growth in
\RNexpMaj{} percent of episodes. If $R$ measures dependence, that
demand should also shape how service output moves year by year. I
split the real growth of market-service value added in each economy
and year, 1996--2019, into the contributions of the four supporting
categories,
\[
g_t=\sum_{k\in\mathcal K} c_{kt},\qquad
c_{kt}=\frac{\widetilde V^S_{k,t}-V^S_{k,t-1}}{V^S_{t-1}},
\]
where $\widetilde V^S_{k,t}$ is the market-service value added that
category $k$ supports in year $t$, valued at year $t-1$ prices. The
share of the variance of $g$ that comes from category $k$ is
$\beta_k=\operatorname{cov}(c_k,g)/\operatorname{var}(g)$. The
$\beta_k$ add to one, as the level shares $V^S_{k,t-1}/V^S_{t-1}$ do,
so the two can be compared directly. If every category grew at the
same rate each year, each variance share would be close to its level
share. A category whose own growth swings more than the total, in step
with it, carries a larger share of the variance.

\begin{figure}[htbp]
\centering
\includegraphics[width=0.52\textwidth]{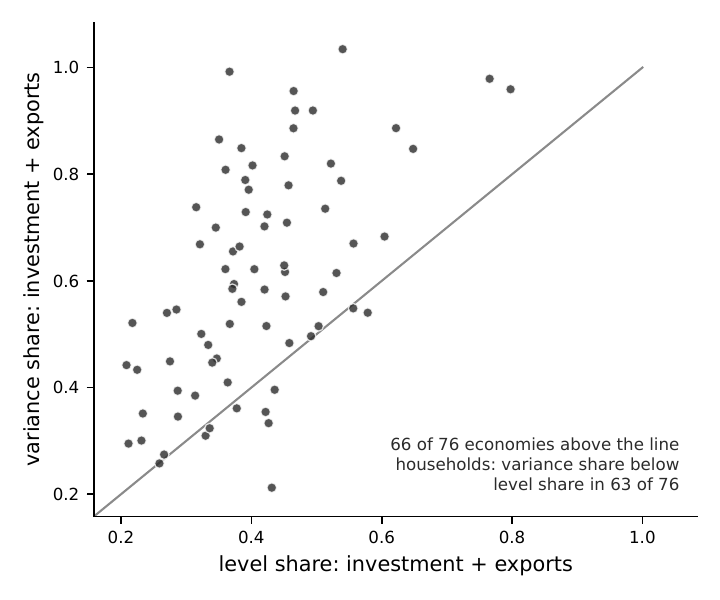}
\caption{Investment and exports carry more of the variance of service
growth than of its level. Each point is an economy, 1996--2019: the
share of market-service value added supported by investment and
exports (horizontal) against their share of the variance of its real
growth (vertical). Points above the 45-degree line: 66 of 76.}
\label{fig:exposure}
\end{figure}

Investment and exports support a median 39 percent of market-service
value added but carry 59 percent of the variance of its real growth.
Their variance share exceeds their level share in 66 of 76 economies
(Figure~\ref{fig:exposure} and Panel~A of Table~\ref{tab:exposure}).
Households support 52 percent and carry 39 percent. The reason is that
household-supported service value added grows most steadily. Its median
standard deviation is 3.7 percent a year, against 7.0 percent for
export-supported and 10.8 percent for investment-supported value added.
It is the least volatile of the four in 71 of 76 economies. Investment
includes changes in inventories, the most volatile item of final
demand. But they support only 0.1 percent of market-service value added
and carry 3 percent of the investment-supported variance share. So the
result is one about fixed capital formation. The pattern
holds in 1996--2007 and in 2010--2019, with variance shares of 0.55 and
0.68 against level shares of 0.39 and 0.41. It also holds per capita
and among economies with more than 5 million people.

This changes one reading of the rise of services. Services are often
treated as the stable sector, and the shift toward them as a source of
lower aggregate volatility \citep{moro2012structural,
moro2015structural}. \citet{carvalho2013great} tie aggregate volatility
to the sectoral composition of output, and
\citet{mirandapinto2021production} finds that denser production
networks lower volatility, more so where the service share is high.
These papers work through sector productivity shocks and production
linkages. \citet{vomlehn2022investment} find that investment goods
come mostly from a few hubs that are more cyclical than other
sectors.

Part of the stability of services also comes from whom they serve. Real market-service value added grows more
steadily than goods value added in 59 of 76 economies (median standard
deviations of 3.5 and 4.4 percent). An accounting counterfactual
measures how much of this gap is the \textit{demand mix}, the shares of
a sector's value added that the four categories support. The
counterfactual gives goods the services' demand mix and keeps each
category's own goods growth. It lowers the median standard deviation of
goods growth to 3.9 percent and closes a median 37 percent of the gap
among those 59 economies. The spread is wide. The interquartile range
runs from $-1$ to 57 percent, the counterfactual widens the gap in 15
of the 59, and pooled across economies it closes 17 percent. Giving services the goods' mix opens a
median 65 percent of the gap (60 percent pooled). Because each
category's own growth contains its productivity shocks, this channel
complements the supply-side account rather than replacing it.

\begin{table}[htbp]
\centering
\begin{threeparttable}
\caption{Whose demand moves services}
\label{tab:exposure}
\footnotesize
\setlength{\tabcolsep}{3pt}
\begin{tabular}{lccccc}
\toprule
\multicolumn{6}{l}{\textit{A. Where the variance of service growth comes from, 1996--2019}} \\
 & Households & Government & Investment & Exports & Inv.\ + exp. \\
\midrule
Level share, median & 0.52 & 0.07 & 0.11 & 0.27 & 0.39 \\
\quad of which inventories & & & 0.00 & & \\
Variance share, median & 0.39 & 0.03 & 0.24 & 0.35 & 0.59 \\
\quad of which inventories & & & 0.01 & & \\
Variance share, pooled & 0.42 & 0.04 & 0.22 & 0.32 & 0.54 \\
\quad of which inventories & & & 0.01 & & \\
Variance share $>$ level share & 13 & 11 & 67 & 52 & 66 \\
\midrule
\multicolumn{6}{l}{\textit{B. Cross-section of economies, 1996--2019}} \\
 & \multicolumn{2}{c}{Log s.d., services} & \multicolumn{2}{c}{Log s.d.\ ratio, services/goods} \\
\cmidrule(lr){2-3}\cmidrule(lr){4-5}
 & (1) & (2) & (3) & (4) \\
\midrule
Domestic household share $R$ & $-$1.20 &  & 0.02 &  \\
 & (0.46) &  & (0.40) &  \\
Export-supported share &  & 1.24 &  & 0.11 \\
 &  & (0.40) &  & (0.39) \\
Investment-supported share &  & 0.32 &  & 1.33 \\
 &  & (1.42) &  & (1.14) \\
Market-service share $s^S$ & $-$2.26 & $-$2.54 & $-$1.61 & $-$1.31 \\
 & (0.70) & (0.79) & (0.62) & (0.60) \\
Log GDP per capita (PPP) & $-$0.12 & $-$0.09 & $-$0.05 & $-$0.08 \\
 & (0.08) & (0.08) & (0.07) & (0.07) \\
Economies & 75 & 75 & 75 & 75 \\
$R^2$ & 0.28 & 0.31 & 0.21 & 0.23 \\
\midrule
\multicolumn{6}{l}{\textit{C. Episodes: median contribution to service growth (points)}} \\
 & Households & Government & Investment & Exports \\
\midrule
2009 (median growth $-$1.8) & 1.0 & 0.4 & $-$1.0 & $-$2.0 \\
2020 (median growth $-$4.6) & $-$1.1 & 0.1 & $-$0.5 & $-$3.2 \\
2009: households up, exports down & \multicolumn{4}{c}{46 of 76 economies} \\
2020: households down, exports up & \multicolumn{4}{c}{6 of 76 economies} \\
\bottomrule
\end{tabular}
\begin{tablenotes}
\footnotesize
\item Real growth of market-service value added, chain-linked from the
previous-year-price tables, split into the contributions of the four
supporting categories. Panel A: level share, each category's share of
market-service value added averaged over 1995--2018; variance share,
$\operatorname{cov}(c_k,g)/\operatorname{var}(g)$ over 1996--2019 in
each economy; medians across 76 economies, and the pooled ratio of
summed covariances to summed variances. The rows ``of which
inventories'' trace the part of investment that is changes in
inventories, from the same tables with fixed capital formation and
inventories kept apart. Panel B: cross-section of 75
economies (Taiwan lacks income data); dependent variable the log
standard deviation of real service growth, or its ratio to that of
goods (industries A--F); shares and income averaged over the window;
robust standard errors. Panel C: medians across 76 economies, in
percentage points of the previous year's market-service value added.
\textit{Source}: OECD inter-country input-output tables in current and
previous-year prices, 2026 release.
\end{tablenotes}
\end{threeparttable}
\end{table}

The 2009 collapse shows the demand mix at work (Panel~C). In the median
economy, real market-service value added fell 1.8 percent.
Household-supported service value added contributed 1.0 point of
growth and government 0.4, while investment subtracted 1.0 and exports
2.0. In 46 of 76 economies, household-supported service value added
rose while export-supported value added fell, and exports were the
largest drag in 49. The 2008--09 collapse in trade was concentrated in durable
and investment goods \citep{bems2010demand,eaton2016trade}. But the
export drag on service value added came more through exports of
services themselves (a median $-1.05$ points) than through services
embodied in goods exports ($-0.66$).
\citet{ariu2016crisis} finds that Belgian exports of business,
telecommunication, and financial services kept growing in 2008--09.
Across these 76 economies, service value added supported by service
exports fell in 60. The year 2020 was different. The median export
contribution, $-3.2$ points, was larger than the household one,
$-1.1$. About half of the export fall reflects spending by foreign
visitors, which counts as exports, so lockdowns abroad appear as an
export shock.

Across economies the evidence is weaker (Panel~B). Without controls,
$R$ is unrelated to the volatility of service growth ($-0.18$, s.e.\
0.43). Holding the service share and income fixed, a 10-point higher
$R$ goes with an 11 percent lower standard deviation of real service
growth ($-1.20$, s.e.\ 0.46). The export-supported share carries the
association. About half of
that coefficient is the composition effect already in Panel~A. The
association halves with a control for population and disappears among
economies with more than 5 million people, so it is mainly a pattern of
small, open economies. The 2008 export share does not predict the size
of the 2009 fall across economies. Nor do economies with a higher $R$
have services that are more stable relative to goods. A shift-share
design built on foreign demand shocks also gives no usable answer
(Appendix~\ref{app:audit}). The robust consequence is the
within-economy one: in the median economy, in every window, investment
and exports carry more of the variance of service growth than of its
level. It follows, as an accounting implication, that moving support
toward them, as 1995--2008 did, ties service growth more closely to the
most volatile parts of final demand.

\subsection{What the service share says about household demand}\label{sec:engel}

Whose demand supports services also changes what the service share can
teach about household demand. In a closed economy with one type of
final demand, domestic households support every unit of value added.
Then the service share equals the service content of household demand,
$s^S=\sigma_C$. In the data the two differ, because
$s^S=\sigma_C\alpha_C/R$. \citet{herrendorf2013two} make the underlying
point for the United States. Sectoral value added includes production
for investment, so it is the wrong object for preferences. They build
consumption value added, whose household-only counterpart $\sigma_C$
measures here. Their comparison is between final expenditure and
consumption value added; the one here is between total value added and
consumption value added.
This subsection measures how far apart the two objects' income
gradients are across 75 economies.

I estimate a log-odds relation of the nonhomothetic CES form
\citep{comin2021structural} on both objects. For a share $w$ of service
value added,
\[
\log\frac{w_{ct}}{1-w_{ct}}
=\gamma\log\frac{p^S_{ct}}{p^N_{ct}}+\beta\log y_{ct}
+\mu_c+\tau_t+u_{ct}.
\]
Here $\mu_c$ is a country effect and $\tau_t$ a year effect. For
$w=s^S$, the price is the chain-linked value-added price of market
services relative to all other industries. For $w=\sigma_C$, both the
share and the relative price refer to the value added that household
consumption supports. Both price indexes come from the current- and
previous-year-price inter-country tables of Section~\ref{sec:prices}.
Income $y$ is GDP per capita at PPP. The relation is a reduced form.
\citet{comin2021structural} estimate their demand system on household
budget data and, across countries, mainly on sectoral employment
shares. Their income variable is real consumption expenditure, with a
base-sector share term and controls for trade. So
$\beta$ here is an income gradient, not their nonhomotheticity
parameter, and the comparison is between the two objects, not with
their estimates.

\begin{table}[htbp]
\centering
\begin{threeparttable}
\caption{Income gradients of the service share and of the service
content of household demand}
\label{tab:inference}
\small
\begin{tabular}{lcccc}
\toprule
 & \multicolumn{2}{c}{Country effects} & \multicolumn{2}{c}{Country and year effects} \\
\cmidrule(lr){2-3}\cmidrule(lr){4-5}
Share in the log odds & Income & Relative price & Income & Relative price \\
\midrule
Service share, $s^S$ & $+0.221$ & $+0.081$ & $-0.014$ & $+0.233$ \\
 & (0.035) & (0.048) & (0.045) & (0.055) \\
Service content of household demand, $\sigma_C$ & $+0.387$ & $+0.024$ & $+0.122$ & $+0.158$ \\
 & (0.046) & (0.080) & (0.072) & (0.087) \\
\midrule
Difference, $\sigma_C$ minus $s^S$ & $+0.166$ & & $+0.136$ & \\
 & (0.037) & & (0.062) & \\
\bottomrule
\end{tabular}
\begin{tablenotes}
\footnotesize
\item Dependent variable: log odds of the share named in the row.
``Income'' is the coefficient $\beta$ on log GDP per capita (PPP);
``Relative price'' is $\gamma$, the coefficient on the log
chain-linked relative value-added price that matches the share. The
difference row stacks the two equations, with object-specific fixed
effects and price slopes. Annual panel, 1995--2018, 75 economies,
1,800 country-years. Standard errors clustered by economy in
parentheses. \textit{Source}: OECD inter-country input-output tables
in current and previous-year prices, 2026 release; World Development
Indicators.
\end{tablenotes}
\end{threeparttable}
\end{table}

The service share rises less with income than the service content of
household demand (Table~\ref{tab:inference}). With country effects, the
income coefficients are 0.22 and 0.39, and the difference is 0.17 (s.e.\
0.04). With common year effects removed as well, the coefficient on the
service share is zero ($-0.01$, s.e.\ 0.045). The one on the service
content of household demand is 0.12 (s.e.\ 0.07). The difference, 0.14
(s.e.\ 0.06), is still significant at the 5 percent level.

The gap is the identity at work. Because $s^S=\sigma_C\alpha_C/R$ and
$\alpha_C$ falls with income faster than $R$ does
(Table~\ref{tab:determinants}), the service share rises less with
income than $\sigma_C$. Part of the gap in log odds is also mechanical.
A share that sits lower has a flatter log-odds slope even when two
shares move in proportion, and $s^S$ sits below $\sigma_C$. For
example, let both shares rise by 1 percent. The log odds then rise by
0.020 at a share of 0.50 but by 0.030 at 0.67, near the world medians of
$s^S$ and $\sigma_C$. So the size of the gap depends on the metric and
on the fixed effects, and I do not read it as a structural parameter.
Its direction is robust. Models in which all final demand is household
consumption, such as \citet{buera2012rise}, equate the service share
with household demand. Matched to the service share's income gradient,
such a model assigns household service demand a weaker income effect
than $\sigma_C$ itself shows. The
price coefficients are positive in every case, consistent with gross
complementarity. But they imply elasticities of substitution close to
one, and the reduced form cannot separate the substitution effect from
the income effect.

\section{Conclusion}\label{sec:conclusion}

The service share is the standard statistic of structural change. This
paper gives it a complement: the part of a country's service value
added that its own households support. The measure traces value added
to final demand for 76 economies, first at home and then across
borders. Five results follow.

First, the service share does not reveal whose demand supports
services. Economies with nearly the same share differ widely in the
part their domestic households support. The difference sits in the
household share of value added, not in what households buy.

Second, domestic households came to support less of their countries'
services: in 61 of 76 economies over 1995--2018, and in 77 percent of
those whose service share rose. Most of the fall came before 2008, and
economies with faster income growth saw larger falls. The service
content of household demand rose, as the Engel pattern predicts, but
the household share of value added fell faster.

Third, the support moved to different places. Exports moved it
abroad, about equally to foreign households and to foreign investment.
Counting households anywhere halves the fall, and households at home
and abroad still support most of the services of 75 of 76 economies.
Within countries, support moved from households to investment as
income grew.

Fourth, whose demand supports services matters. Services inherit the
volatility of the demand behind them. Investment and exports carry more
of the variance of service growth than of its level, and part of why
services are smoother than goods is whom they serve. In 2009,
household-supported service value added grew in the median economy
while export-supported value added fell. The income effect in household
service demand also looks weaker when read through the service share
than through the service content of household demand.

Fifth, an exact accounting of the changes allocates most of the growth
in the service share to the production system. That group holds the
input and value-added coefficients and the industry composition of
investment and export demand. Almost all of the fall in the domestic
household share goes to the shares of the four final users in total
final demand. The composition of household demand raised the
domestic household share where it moved; the falling weight of
households overwhelmed it.

One caution concerns prices. The main counts and the timing of the fall
hold in chain volume, and the domestic household share depends less on
the price basis than the service share does. But China's real
service-share trend depends on how industry value added is deflated.

The policy reading follows. A government that sets a target for the
service share can meet it without any change in household consumption.
Higher relative service prices can do it; they account for most or all
of China's nominal increase, depending on the deflator. So can
investment and export demand that reaches service producers through
supply chains. A service sector built that way also carries more of the
volatility of investment and foreign demand. The domestic household
share is unchanged by a common revaluation of service value added and rises
only when domestic households support a larger part of service
production. Where the aim is a shift toward domestic consumption, it is
the statistic to watch.

\section*{Data Availability Statement}

All primary data sources are publicly available. The OECD publishes the
harmonized national input-output tables and the inter-country
input-output tables in current and previous-year prices at
\texttt{oecd.org}. It also publishes the annual non-financial
institutional-sector accounts (P3, P31, and B6G). The other source is
the World Bank World Development Indicators (retrieved via the public
API). China's official
national input-output tables come from the National Bureau of
Statistics public data portal (\texttt{data.stats.gov.cn}). The
complete $R$/$H$ dataset is released with the submission. It holds
$R_{ct}$ with its components $\alpha_C$, $\sigma_C$ and $s^S$ for 76
economies over 1995--2022, and $H_{ce}$ with the full final-demand
composition of every episode. The households-anywhere shares, the
chain-volume series, and $H$ in national currency come with it. The
construction and analysis code is available from the author during
review and will be deposited in a permanent public archive upon
publication. It rebuilds the dataset from the national tables and
regenerates the exhibits that use those tables alone, with a list of
each exhibit's source. The exhibits built from the inter-country and
previous-year-price tables have their own programs, supplied with the
code.

\appendix

\section{Construction Details}\label{app:construction}

\textit{Source}: OECD harmonized national input-output tables, 2025
release, 50 ISIC Rev.\ 4 industries, 76 economies, 1995--2022 (2,128
country-year tables). They are domestic use tables with import matrices
(domestic plus imported product rows, basic prices), with value-added
and gross-output rows.

\textit{Direct shares}: total use $=$ intermediate use $+$ HFCE $+$
NPISH $+$ government $+$ GFCF $+$ inventories $+$ residents' direct
purchases abroad $+$ non-resident purchases $+$ exports. The imports
adjustment column is excluded. Households are grouped by residence:
the household share counts HFCE plus residents' direct purchases
abroad, which appear only in the import rows. The China comparison with
the official tables (Appendix~\ref{app:validation}) reports the
household share both with and without them.

\textit{Traced shares}: For category $k$, gross output is
$x^k=(I_n-A)^{-1}f^k=Lf^k$. The domestic final-demand vectors cover
household consumption (HFCE), government (NPISH$+$GGFC), investment
(GFCF$+$inventories), and exports (exports$+$non-resident purchases).
$\sum_k x^k$ reproduces gross output to rounding in every table, and
industry value-added coefficients map $x^k$ into value added.

\textit{Groups}: consumer-facing $=$ \{G, I, R, S, T\};
producer $=$ \{H49--H53, J58T60, J61, J62\_63, M, N\}; FIRE $=$ \{K,
L\}. Market services are their union; O, P, and Q are excluded.

\textit{Consumer-facing measures}: the consumer-facing robustness
changes only the producing-sector selector. Let $e_{CF}$ select trade,
accommodation and food, recreation, and personal services. Then
\[
 V_k^{CF}=e_{CF}'\widehat v Lf^k,\qquad
 R^{CF}=\frac{V_C^{CF}}{V^{CF}},\qquad
 H^{CF}=\frac{\Delta V_C^{CF}}{\Delta V^{CF}}.
\]
Its episodes are selected using the consumer-facing value-added share
and consumer-facing value-added growth.

\textit{$R$ and $H$}: $R_{ct}$ $=$ household-supported market-service
VA over market-service VA. The country-windows are 1995--2005,
2005--2015, and 2015--2022. A country-window is a tertiarization
episode if its market-service VA share rises and its nominal
market-service VA is positive and increasing over the same window.
$H_{ce}$ $=$ household contribution to the service-VA
expansion; all episodes have strictly positive service-VA growth. The
national-currency $H$ converts both endpoints with the World
Development Indicators conversion factor implied by services value
added in current local currency and in current U.S. dollars. The factor
is one for the United States. It covers \RNHlcuN{} of the 161 episodes;
Taiwan, and Canada and Hong Kong in 1995--2005 and Myanmar in
2005--2015, lack a factor at one endpoint. \textit{Direct--traced
comparisons} (Figure~\ref{fig:ultimate}) use domestic flows on both
sides.

\textit{Exact decomposition}: $\alpha_C$ $=$ household-supported total VA
over total VA; $\sigma_C$ $=$ household-supported market-service VA over
household-supported total VA; $s^S$ $=$ market-service VA over total
VA. The identity $R = \alpha_C \sigma_C/s^S$ holds to machine precision
in all 2,128 tables.

\textit{Household service budget}: For the supplementary
disposable-income measure, $E_C^S/E_C$ is the direct market-service
share of resident household and NPISH consumption. The input-output
construction includes HFCE, NPISH, and residents' direct purchases
abroad and excludes non-residents' purchases. It is multiplied by
$P3(S1M)/B6G(S1M)$ from the OECD institutional-sector accounts
\citep{oecd2026nasec}. Here S1M denotes households plus NPISH, P3 final
consumption expenditure, and B6G gross disposable income, so that
\[
\frac{E_C^S}{Y_C^{\mathrm{disp}}}
=
\frac{E_C^S}{E_C}\frac{E_C}{Y_C^{\mathrm{disp}}},
\]
with $E_C$ resident household and NPISH consumption expenditure,
$E_C^S$ its direct expenditure on market services, and
$Y_C^{\mathrm{disp}}$ gross disposable income. The resulting
$E_C^S/Y_C^{\mathrm{disp}}$ measure covers 42 economies in 2018 and 1,085
country-years over 1995--2022. Using household-sector P31 and B6G with
HFCE and direct purchases abroad gives the household-only robustness
measure for 36 economies in 2018. It is nearly identical (correlation
0.993; mean absolute difference 0.006). The median ratio in 2018 is
0.54 and China's is 0.31. The service share of household consumption
rises with income (0.045, robust s.e.\ 0.019); after scaling by
disposable income the gradient is positive but imprecise (0.037,
robust s.e.\ 0.032). The ratio's 2018 correlation with $R$ is only
0.11, against 0.67 for $\sigma_C$. Household budgets tilt toward
services as income rises, but that does not identify who supports the
service sector. India is not covered by the OECD sector accounts.

\textit{Price decomposition}: WDI current- and constant-LCU value
added for services and industry give sectoral implicit deflators. The
two-sector nominal share is services over services plus industry, and
the real share values constant-price quantities at base-year prices.
The China split (15.9 points nominal $=$ 1.1 real $+$ 14.8 relative
price) is identical to one decimal under 1995-base (Laspeyres-type)
and 2018-base (Paasche-type) valuation. So the base year and
chain-link nonadditivity do not affect this two-sector split. The
comparators are India, $+7.8$ nominal $=$ $+6.6$ real $+$ $1.1$ price,
and Korea, $+3.6$ $=$ $-1.1$ real $+$ $4.7$ price. The WDI services
aggregate is broader than the paper's market-service definition, since
it includes public administration, education, and health. So the
decomposition serves only as a comparable two-sector price diagnostic
and is never used to re-measure $R$. Section~\ref{sec:prices} replaces it with
chain-volume results from the inter-country tables in previous-year
prices.

\textit{Data release}: the complete $R$/$H$ dataset (version 3, built
on the 2025 release) accompanies the paper. It contains $R_{ct}$ with
its components $(\alpha_C, \sigma_C, s^S)$ for 76 economies $\times$
1995--2022 (2,128 country-years) and $H_{ce}$ with the full
final-demand composition for all 161 tertiarization episodes. Value
added by supporting category, the shift-share panel, and the
shift-share exposures come with it. The accompanying code archive
rebuilds the files of the release that come from the national tables.
It regenerates the exhibits that use those tables alone. The exhibits
built from the inter-country and previous-year-price tables have their
own programs, supplied with the code. A rest-of-world aggregate
(``ROW'') in the source is excluded throughout. So are the four
economies outside the fixed analysis sample (Angola, the United Arab
Emirates, the Democratic Republic of the Congo, and S\~ao Tom\'e and
Pr\'incipe).

\section{Direct-Purchaser Patterns}\label{app:direct}

This appendix reports the direct layer: who hands over the money. It
is the measurement floor under the paper's final-demand analysis.
Table~\ref{tab:levels} gives the direct shares by group over 1995--2022,
and Table~\ref{tab:fractions} shows how general the firm tilt is. At
the world median, intermediate users buy 36 percent of consumer-facing
service output. Intermediate use exceeds household purchases of market
services in 84 percent of economies.

\begin{figure}[H]
\centering
\includegraphics[width=0.58\textwidth]{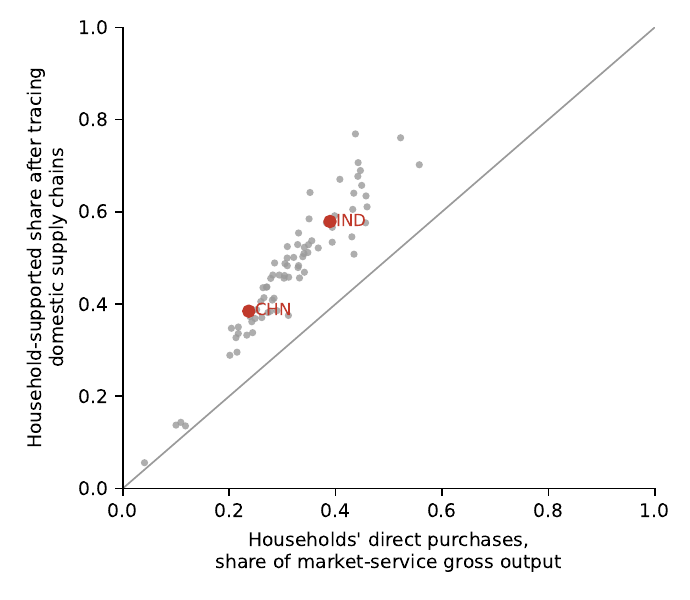}
\caption{Direct and traced household shares of market-service gross
output, 76 economies, 2018. The vertical gap is market-service output
first sold to firms but required by household consumption once
domestic supply chains are traced. Tracing can only add to the direct
share, so every economy lies above the 45-degree line (median gap: 15
points). China and India are labeled in
red (China: 0.24 direct, 0.38 traced; India: 0.39 direct, 0.58
traced).}
\label{fig:ultimate}
\end{figure}

Figure~\ref{fig:ultimate} isolates the tracing step. Let $f=\sum_k
f^k$. The x-axis is $D_C=e_S'f^C/e_S'Lf$: households' direct final
purchases from market-service industries as a share of those
industries' gross output. The y-axis is $U_C=e_S'Lf^C/e_S'Lf$:
market-service gross output ultimately supported by household final
demand after tracing all domestic input linkages. Both axes have the
same denominator, so $U_C-D_C$ is market-service output sold at first
to firms but ultimately required by household consumption. Tracing can
only add, since $U_C\geq D_C$ by construction. These gross-output
shares diagnose the tracing step; they are not the paper's value-added
measure $R$.

\begin{table}[H]
\centering
\begin{threeparttable}
\caption{Direct purchaser shares: world medians and China, 1995--2022}
\label{tab:levels}
\small
\begin{tabular}{llcccccc}
\toprule
Group & Share & 1995 & 2005 & 2010 & 2015 & 2018 & 2022 \\
\midrule
Market services & median intermediate & 0.446 & 0.463 & 0.468 & 0.460 & 0.455 & 0.452 \\
                & median household    & 0.347 & 0.325 & 0.316 & 0.317 & 0.314 & 0.306 \\
                & China: intermediate & 0.558 & 0.612 & 0.571 & 0.639 & 0.591 & 0.606 \\
                & China: household    & 0.249 & 0.213 & 0.224 & 0.211 & 0.242 & 0.228 \\
\midrule
Consumer-facing & median intermediate & 0.371 & 0.371 & 0.386 & 0.366 & 0.364 & 0.381 \\
                & median household    & 0.387 & 0.371 & 0.382 & 0.384 & 0.384 & 0.360 \\
                & China: intermediate & 0.521 & 0.576 & 0.542 & 0.607 & 0.564 & 0.597 \\
                & China: household    & 0.295 & 0.234 & 0.251 & 0.238 & 0.271 & 0.246 \\
\midrule
Producer        & median intermediate & 0.495 & 0.510 & 0.519 & 0.516 & 0.507 & 0.501 \\
                & median household    & 0.323 & 0.301 & 0.291 & 0.286 & 0.277 & 0.276 \\
                & China: intermediate & 0.582 & 0.633 & 0.587 & 0.655 & 0.605 & 0.610 \\
                & China: household    & 0.219 & 0.201 & 0.210 & 0.197 & 0.227 & 0.218 \\
\bottomrule
\end{tabular}
\begin{tablenotes}
\footnotesize
\item Shares of total use (intermediate plus final demand). Households
include residents' direct purchases abroad. $n = 76$ economies each year.
\end{tablenotes}
\end{threeparttable}
\end{table}

\begin{table}[H]
\centering
\begin{threeparttable}
\caption{How general is the direct firm tilt? Cross-country fractions}
\label{tab:fractions}
\small
\begin{tabular}{lcccc}
\toprule
& \multicolumn{2}{c}{Intermediate $>$ household} &
  \multicolumn{2}{c}{Intermediate $>$ 1/2 of use} \\
\cmidrule(lr){2-3}\cmidrule(lr){4-5}
Group & 2005 & 2018 & 2005 & 2018 \\
\midrule
Market services & 87\% & 84\% & 22\% & 17\% \\
Consumer-facing & 46\% & 42\% & \phantom{0}3\% & \phantom{0}4\% \\
Producer        & 89\% & 91\% & 57\% & 54\% \\
\bottomrule
\end{tabular}
\begin{tablenotes}
\footnotesize
\item Fraction of 76 economies. An intermediate share above one-half is
equivalent to intermediate use exceeding all final-demand categories
combined; in market services intermediate use is the \textit{largest}
purchaser category nearly everywhere but a majority only in a small
minority of economies. In consumer-facing services households outbuy
intermediate users in a small majority of economies, yet intermediate
users still take about a third of direct use at the median, so using
consumer-facing sector labels as proxies for household demand
misclassifies roughly a third of that use.
\end{tablenotes}
\end{threeparttable}
\end{table}

\begin{table}[H]
\centering
\begin{threeparttable}
\caption{Sector detail, 2015: world median versus China (direct shares)}
\label{tab:sectors}
\small
\begin{tabular}{lcccc}
\toprule
& \multicolumn{2}{c}{World median} & \multicolumn{2}{c}{China} \\
\cmidrule(lr){2-3}\cmidrule(lr){4-5}
Sector & Intermediate & Household & Intermediate & Household \\
\midrule
Wholesale/retail (G) & 0.454 & 0.306 & 0.706 & 0.139 \\
Accommodation/food (I) & 0.145 & 0.660 & 0.411 & 0.569 \\
Arts/recreation (R) & 0.136 & 0.485 & 0.120 & 0.150 \\
Other services (S) & 0.231 & 0.473 & 0.413 & 0.378 \\
Land transport (H49) & 0.499 & 0.245 & 0.787 & 0.101 \\
Telecoms (J61) & 0.434 & 0.422 & 0.651 & 0.298 \\
Professional (M) & 0.641 & 0.041 & 0.776 & 0.014 \\
Finance (K) & 0.593 & 0.279 & 0.756 & 0.182 \\
Real estate (L) & 0.213 & 0.735 & 0.240 & 0.669 \\
\bottomrule
\end{tabular}
\begin{tablenotes}
\footnotesize
\item Shares of total use; households include residents' direct
purchases abroad. Worldwide, hotels and restaurants sell
mostly to households; on the harmonized tables China's
intermediate share (0.41) is nearly three times the world median,
though below its household share (0.57), and China's own official tables put the
intermediate share near or above one-half, exceeding it in 2015 and
2017 (Appendix~\ref{app:validation}). The tilt is consistent with a large
role for business meals and official entertainment, though the use
table records the purchaser, not the occasion.
\end{tablenotes}
\end{threeparttable}
\end{table}

\section{China and India, Component by Component}
\label{app:chinaext}

This appendix gives the China--India comparison component by
component.

\begin{table}[H]
\centering
\begin{threeparttable}
\caption{The China--India pair, component by component (2018)}
\label{tab:chnind}
\small
\begin{tabular}{lccc}
\toprule
Component & China & India & World median \\
\midrule
$\alpha_C$: household share of value added & 0.324 & 0.518 & 0.376 \\
$\sigma_C$: service content of household demand & 0.539 & 0.454 & 0.669 \\
$s^S$: service share & 0.415 & 0.402 & 0.497 \\
$R = \alpha_C \sigma_C / s^S$: domestic household share of services & 0.421 & 0.585 & 0.495 \\
\bottomrule
\end{tabular}
\begin{tablenotes}
\footnotesize
\item Similar service shares, different domestic household shares; the
divergence sits almost entirely in the household share of value added
($\alpha_C$), not in the service content of household demand
($\sigma_C$), which is modest in both. The identity holds
exactly within each economy (China: $0.324\times0.539/0.415=0.421$;
India: $0.518\times0.454/0.402=0.585$). It does not hold across the
world-median column, whose entries are separate medians of each
component and so need not multiply to the median $R$.
\end{tablenotes}
\end{threeparttable}
\end{table}

\section{Validating the Chinese Cells Against the Official Tables}
\label{app:validation}

China's accommodation-and-food inversion is extreme enough to warrant a
direct check against the primary source. I extracted the final-use
blocks of China's official national input-output tables from the
National Bureau of Statistics database. They are the 2012 benchmark
(139 sectors), the 2015 extension (42 sectors), and the 2017 benchmark
(149 sectors). I aggregated the accommodation and catering rows and
computed the same purchaser shares as in the harmonized data.

\begin{table}[H]
\centering
\begin{threeparttable}
\caption{Accommodation and food in China: official tables versus
OECD-harmonized tables}
\label{tab:validation}
\small
\begin{tabular}{lccccc}
\toprule
& \multicolumn{2}{c}{Official NBS tables} &
  \multicolumn{3}{c}{OECD-harmonized} \\
\cmidrule(lr){2-3}\cmidrule(lr){4-6}
& & & & \multicolumn{2}{c}{Household} \\
\cmidrule(lr){5-6}
Year & Intermediate & Household & Intermediate & HFCE only & With purchases abroad \\
\midrule
2012 & 0.498 & 0.479 & 0.408 & 0.502 & 0.566 \\
2015 & 0.605 & 0.378 & 0.411 & 0.443 & 0.569 \\
2017 & 0.578 & 0.411 & 0.407 & 0.472 & 0.579 \\
\bottomrule
\end{tabular}
\begin{tablenotes}
\footnotesize
\item Shares of total use (imports excluded), accommodation $+$ catering
rows (sector 31 in the 42-sector 2015 table; name-matched rows in the
139/149-sector benchmarks) versus sector I of the OECD tables. The
official tables have no column for residents' direct purchases abroad,
so the like-for-like OECD household share counts HFCE only; the last
column adds those purchases, the residence basis of
Appendix~\ref{app:direct}.
Alternative denominators move the shares by 1--6 points: on domestic
absorption the official household share is 0.385--0.491; crediting all
non-resident purchases to households raises the HFCE-only share by
1.3--2.5 points (non-resident spending is negligible for China, ruling
out the tourism-classification explanation). OECD-harmonized columns
are on the 2025 release, whose revision lowered the harmonized
intermediate shares (0.38 in 2018), widening the official--harmonized
gap: the majority-intermediate characterization rests on the official
tables, and the harmonized figures are the conservative floor.
\end{tablenotes}
\end{threeparttable}
\end{table}

The official tables put the intermediate share of Chinese accommodation
and food \textit{higher} than the harmonized tables in every available
year, so the paper's main sector comparison is conservative. On the
like-for-like basis, without residents' purchases abroad, the household
shares differ by 2 to 7 points. Adding those purchases, as
Appendix~\ref{app:direct} does, raises the harmonized household share
to 0.57--0.58. The reason is that Chinese residents' accommodation and
food bought abroad equals 6--13 percent of total use in these years.
The result is insensitive to the choice of denominator. The
use table cannot resolve the \textit{occasion} behind the purchaser.
Business meals, official entertainment, institutional catering, and
platform-intermediated delivery are all consistent with the split. So
the text attributes the inversion to intermediate purchasers, not to
any single practice.

\section{Definition and Window Robustness}
\label{app:definitions}

\subsection{Alternative service-sector definitions}

Table~\ref{tab:definitions} repeats the paper's main results under
the paper's definition and four alternatives: two broader ones, one
without finance and real estate, and the consumer-facing subset. The
raw $(s,R)$ correlation is negative under every definition but the
last. It is $+0.14$ for consumer-facing services, with a
$t$-statistic of 1.3, and so supplies no positive level relationship
on which a household-upgrading interpretation can rest.
The income-adjusted partial correlations remain small. More
importantly, the median household contribution to tertiarization never
reaches one-half, the income gradient of $H$ is negative and
significant throughout, and China and India diverge under all five
definitions. For consumer-facing services, median $H^{CF}=0.454$, only
39 percent of 113 episodes are majority-household, and the income
gradient is $-0.147$ ($t=-2.6$). Narrowing the sector changes the
uninformative level correlation but not the flow conclusion. Under the
conventional all-services definition, the household
reading fares worse, not better: the majority-household fraction falls
to 0.24, because education and health are mostly
government-supported. It also refines the result.
Household-\textit{financed} tertiarization is even rarer than
household-supported tertiarization, since part of what households consume
in the broadest service sector is socially financed. Excluding
government changes nothing: $R^{\neg G}=V_C^S/(V_C^S+V_I^S+V_X^S)$
correlates 0.994 with $R$, has partial correlation 0.04 with $s^S$
given income, and keeps China (0.48) and India (0.60) apart.

\begin{table}[H]
\centering
\begin{threeparttable}
\caption{Main results under five service-sector definitions}
\label{tab:definitions}
\scriptsize
\setlength{\tabcolsep}{2.2pt}
\begin{tabular}{lccccccccc}
\toprule
& corr$(s,R)$ & partial & median & frac. & Episodes & income grad. & China & India & gov.\ share \\
Definition & 2018 ($t$) & $|$ income & $H$ & $H{>}0.5$ & $n$ & of $H$ ($t$) & $s$/$R$ & $s$/$R$ & of svc.\ VA \\
\midrule
All services (incl.\ O, P, Q) & $-0.24$ ($-2.1$) & $+0.25$ & 0.386 & 0.24 & 161 & $-0.082$ ($-4.7$) & .52/.36 & .53/.51 & 24\% \\
Market $+$ education/health  & $-0.29$ ($-2.6$) & $+0.16$ & 0.423 & 0.33 & 163 & $-0.091$ ($-4.2$) & .47/.40 & .46/.58 & 16\% \\
Market services (paper)      & $-0.27$ ($-2.4$) & $+0.05$ & 0.453 & 0.41 & 161 & $-0.098$ ($-4.8$) & .41/.42 & .40/.58 & 7\% \\
Market excl.\ FIRE           & $-0.17$ ($-1.5$) & $+0.09$ & 0.335 & 0.25 & 158 & $-0.116$ ($-5.7$) & .27/.34 & .28/.48 & 8\% \\
Consumer-facing services    & $+0.14$ ($+1.3$) & $+0.15$ & 0.454 & 0.39 & 113 & $-0.147$ ($-2.6$) & .14/.38 & .14/.61 & 7\% \\
\bottomrule
\end{tabular}
\begin{tablenotes}
\scriptsize
\item Correlations: 2018 cross-section ($n = 75$ with income data).
Episodes: tertiarization country-windows under each definition's own
service share and value-added growth. The consumer-facing subset has
113 episodes; the broad definitions have 158--163. Income gradient:
$H$ on log GDP pc
at the episode start, window fixed effects, country-clustered. Gov.\
share: world median government-supported share of the definition's
service VA, 2018. Episode windows: 1995--2005,
2005--2015, 2015--2022.
\end{tablenotes}
\end{threeparttable}
\end{table}

\subsection{The correlation by year}

\begin{figure}[H]
\centering
\includegraphics[width=0.74\textwidth]{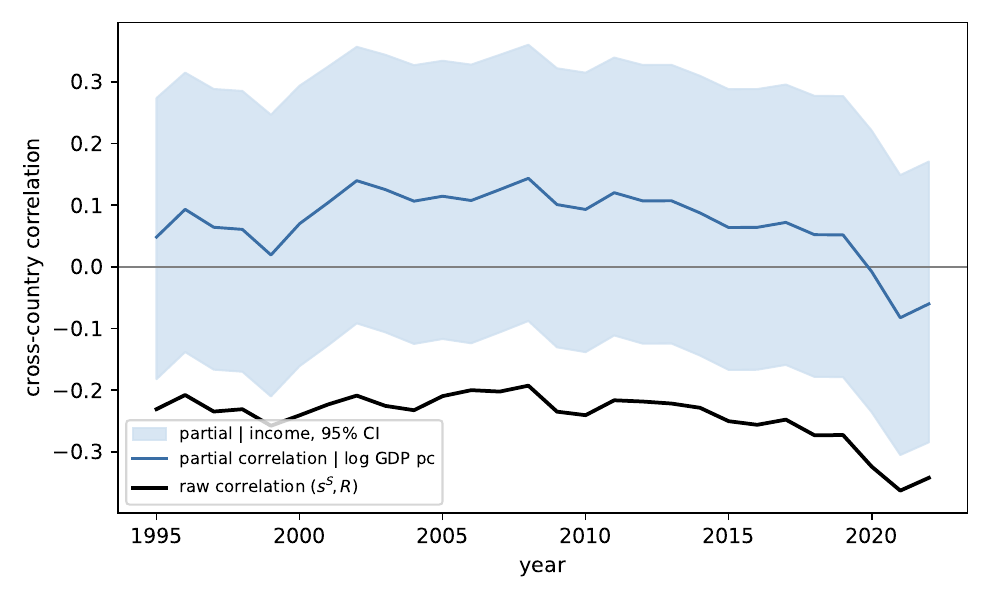}
\caption{The failure is not a single cross-section: the correlation
between the service share $s^S$ and the domestic household share $R$, computed
annually across economies, 1995--2022. The raw correlation (lower
line) is negative in every year; the 95 percent confidence interval of
the income-adjusted partial correlation (upper line, shaded) contains
zero in every year.}
\label{fig:rhotime}
\end{figure}

Figure~\ref{fig:rhotime} repeats the 2018 cross-section of
Section~\ref{sec:reveal} for every year from 1995 to 2022.

\subsection{Alternative tertiarization windows}

Table~\ref{tab:windows} shows the continuous-$H$ results under
alternative window definitions. The median household contribution sits
between 0.44 and 0.48, and the majority-household fraction between 0.40
and 0.46 in every variant. If an episode is defined only by positive
service-VA growth, without requiring a rising service share, the sample
expands to 218 windows. The median $H$ is then 0.44 and the
majority-household fraction 0.39.

\begin{table}[H]
\centering
\begin{threeparttable}
\caption{$H$ under alternative window definitions (market services)}
\label{tab:windows}
\small
\begin{tabular}{lccc}
\toprule
Window definition & Episodes & Median $H$ & Frac.\ $H>0.5$ \\
\midrule
Baseline (1995--2005, 2005--15, 2015--22) & 161 & 0.453 & 0.41 \\
Non-overlapping five-year                 & 242 & 0.444 & 0.40 \\
Rolling five-year                         & 975 & 0.479 & 0.46 \\
Rolling ten-year                          & 995 & 0.474 & 0.44 \\
Baseline, balanced countries              & 161 & 0.453 & 0.41 \\
Baseline, min.\ 10\% service-VA growth    & 157 & 0.453 & 0.41 \\
\bottomrule
\end{tabular}
\begin{tablenotes}
\footnotesize
\item Tertiarization episodes: rising nominal market-service VA share
with positive service-VA growth. Adding the auxiliary requirement of
positive aggregate-VA growth drops two episodes (159) and leaves every
statistic unchanged. The non-overlapping five-year variant appends a
final 2020--2022 window.
\end{tablenotes}
\end{threeparttable}
\end{table}

\section{Chain-Volume Sign Changes}
\label{app:signchanges}

Table~\ref{tab:signchanges} lists the economies whose 1995--2018 change
in the service share or in $R$ differs in sign between current prices
and chain volume. Volumes use 1995 reference prices. Some volume
changes in the service share are very large, above 15 points for
Brunei, Russia, and Norway. Section~\ref{sec:prices} shows for China
how the volume change depends on the deflation of industry value added,
and the paper treats current-price changes as the baseline.

\begin{table}[H]
\centering
\begin{threeparttable}
\caption{Economies whose 1995--2018 changes switch sign in chain volume}
\label{tab:signchanges}
\small
\begin{tabular}{lrrrr}
\toprule
& \multicolumn{2}{c}{$\Delta s^S$ (points)} & \multicolumn{2}{c}{$\Delta R$ (points)} \\
\cmidrule(lr){2-3}\cmidrule(lr){4-5}
Economy & Current & Volume & Current & Volume \\
\midrule
\multicolumn{5}{l}{\textit{Sign of $\Delta s^S$ differs}} \\
BGD & $+0.07$ & $-6.82$ & $-12.75$ & $-22.30$ \\
BLR & $+3.06$ & $-3.34$ & $-8.98$ & $+12.39$ \\
CHN & $+11.92$ & $-6.19$ & $-3.44$ & $-7.68$ \\
JPN & $+2.33$ & $-1.80$ & $-2.75$ & $-1.09$ \\
KOR & $-0.04$ & $+0.20$ & $-6.69$ & $-12.02$ \\
MEX & $-0.59$ & $+7.86$ & $-9.00$ & $-9.04$ \\
MMR & $+1.59$ & $-0.21$ & $-29.74$ & $-34.59$ \\
NOR & $-2.91$ & $+15.09$ & $-2.14$ & $+2.84$ \\
RUS & $-0.27$ & $+20.13$ & $-4.28$ & $-5.69$ \\
SAU & $-1.11$ & $+9.87$ & $-13.61$ & $-14.96$ \\
SVK & $+2.73$ & $-10.23$ & $-5.50$ & $-12.24$ \\
THA & $-0.35$ & $+2.52$ & $-6.50$ & $-7.90$ \\
TUR & $+2.19$ & $-0.72$ & $-9.95$ & $-11.21$ \\
UKR & $+6.19$ & $-2.25$ & $+8.56$ & $+10.79$ \\
VNM & $-1.11$ & $+0.54$ & $-8.56$ & $-17.70$ \\
\midrule
\multicolumn{5}{l}{\textit{Sign of $\Delta R$ differs}} \\
BLR & $+3.06$ & $-3.34$ & $-8.98$ & $+12.39$ \\
BRN & $+1.25$ & $+26.58$ & $+2.35$ & $-0.68$ \\
CMR & $+3.24$ & $+11.61$ & $+1.68$ & $-2.08$ \\
HKG & $+3.83$ & $+6.30$ & $+4.52$ & $-1.21$ \\
IDN & $+2.76$ & $+12.63$ & $+3.34$ & $-0.77$ \\
KAZ & $+5.23$ & $+11.20$ & $-7.92$ & $+2.61$ \\
NGA & $+6.44$ & $+9.21$ & $+22.76$ & $-1.54$ \\
NOR & $-2.91$ & $+15.09$ & $-2.14$ & $+2.84$ \\
PAK & $+1.62$ & $+6.85$ & $+2.88$ & $-1.47$ \\
\bottomrule
\end{tabular}
\begin{tablenotes}
\footnotesize
\item Changes in percentage points. Volume: chain-linked from the
previous-year-price tables with 1995 reference prices, as in
Table~\ref{tab:prices}. \textit{Source}: OECD inter-country
input-output tables in current and previous-year prices, 2026 release.
\end{tablenotes}
\end{threeparttable}
\end{table}

\section{A Foreign-Demand Shock Check}
\label{app:audit}

A shift-share design built on foreign demand shocks gives no usable
answer. I asked whether such shocks, transmitted through each economy's
production network, move service employment or the export-supported
share of service value added. The exposures are 1995--97 network
averages of service value added per unit of final demand for industry
$j$'s output. The shock is the growth of high-income economies' imports
of product $j$, excluding the own country; a variant uses world export
demand excluding the own country. The appropriate inference with 50
(effectively about 13) common shocks uses shock-level aggregation and
industry clustering. With it, the employment coefficient is null with
the wrong sign, and a lead is significant when leads and lags enter
jointly. The export-share estimate is positive but imprecise, its
leave-one-industry-out range crosses zero, and an investment shock
enters with a coefficient that should be zero but is not. So the paper
draws no causal conclusion from this exercise and relies on the
accounting results alone.

\section{Inside the Production Block}
\label{app:productionblock}

Within the production block, most of the allocation goes to the input
mix. Table~\ref{tab:productionblock} splits the block's contribution
into three parts: input use by input, value-added coefficients, and the
domestic share of each input. The split is exact and does not depend on
an order; the method is at the end of this appendix. Among the 56
economies, greater use of market-service inputs receives a median 35
percent of the service-share rise. Its contribution is positive in 48
of them. Lower use of goods inputs adds a median 11 percent. The median
market-service input per dollar of gross output rose from 19 to 24
cents, and the median goods input fell from 23 to 19 cents. Business
services (M and N) are the largest single input: a median 16 percent,
positive in 51 economies. Information and communication adds 3
percent, and finance adds nothing. Value-added coefficients and domestic input shares each
net out near zero, because each combines two moves of opposite sign.
Value added per unit of output fell in goods industries in 43
economies, which raised the service share, and in market services in
36, which lowered it. Imports replaced domestic goods inputs in 51
economies, which raised the service share. They also replaced some
domestic service inputs, which lowered it.

The price basis changes the leading input but not the main result. In
chain volume, market-service inputs still receive the largest part, a
median 21 percent of the rise. Information and communication then lead
with 9 percent, and business services fall to 4 percent. The
difference comes from deflation, not from chaining. Summing
adjacent-year splits at current prices still gives business services
15 percent. From 1995 to 2018, business-service gross-output prices
rose by a median 10 log points relative to the gross-output price of
all industries. Telecommunications prices fell by 63 log points.

\begin{table}[H]
\centering
\begin{threeparttable}
\caption{The production block's contribution to the service-share rise,
by part, 1995--2018}
\label{tab:productionblock}
\scriptsize
\setlength{\tabcolsep}{4pt}
\begin{tabular}{lrrrrrr}
\toprule
&\multicolumn{3}{c}{Current prices}&\multicolumn{3}{c}{Chain volume}\\
\cmidrule(lr){2-4}\cmidrule(lr){5-7}
Part&Median&Pooled&Sign&Median&Pooled&Sign\\
\midrule
\multicolumn{7}{l}{\textit{Total input use, by input}}\\
\quad Market-service inputs&$.35$&$.30$&$.86$&$.21$&$.25$&$.92$\\
\quad \quad Business services (M, N)&$.16$&$.15$&$.91$&$.04$&$.06$&$.63$\\
\quad \quad Information and communication (J)&$.03$&$.04$&$.68$&$.09$&$.09$&$.92$\\
\quad \quad Finance (K)&$-.00$&$-.00$&$.48$&$.02$&$.03$&$.59$\\
\quad \quad Trade (G)&$.05$&$.06$&$.70$&$.05$&$.06$&$.66$\\
\quad \quad Transport (H)&$.03$&$.02$&$.68$&$-.00$&$-.00$&$.49$\\
\quad \quad Other market services (I, L, R--T)&$.03$&$.04$&$.61$&$.00$&$.01$&$.51$\\
\quad Goods inputs (A--C)&$.11$&$.13$&$.84$&$.05$&$.10$&$.81$\\
\quad Utilities, construction, public services (D--F, O--Q)&$-.00$&$-.01$&$.48$&$.01$&$.02$&$.59$\\
Value-added coefficients $v$&$.01$&$.02$&$.50$&$.05$&$.02$&$.54$\\
\quad Goods industries (A--C)&$.14$&$.10$&$.66$&$.05$&$.03$&$.59$\\
\quad Market services&$-.10$&$-.11$&$.39$&$-.09$&$-.09$&$.37$\\
Domestic share of inputs&$.02$&$-.01$&$.57$&$.01$&$.02$&$.61$\\
\quad Market-service inputs&$-.09$&$-.12$&$.21$&$-.06$&$-.06$&$.27$\\
\quad Other inputs&$.10$&$.11$&$.88$&$.07$&$.08$&$.88$\\
\midrule
Production block $(A,v)$&$.48$&$.45$&$.91$&$.42$&$.41$&$.92$\\
Economies&\multicolumn{3}{c}{56}&\multicolumn{3}{c}{59}\\
\bottomrule
\end{tabular}
\begin{tablenotes}\footnotesize
\item Economies with $\Delta s^S>.02$. ``Median'' is the median across
economies of the part's contribution divided by $\Delta s^S$;
``Pooled'' is the sum of the part's contributions over the sum of
$\Delta s^S$; ``Sign'' is the fraction of economies in which the part
has the sign of $\Delta s^S$. Pooled shares add up to the block's
pooled share; medians need not. Input use is the total (domestic plus
imported) input coefficient, by supplying industry. The domestic share
is the share of each input bought from domestic producers. A negative
entry means that imports replacing domestic inputs lowered the service
share. Indented rows under value-added coefficients show two of four
industry groups. Current prices: national tables, one step from 1995
to 2018. Chain volume: inter-country tables, 2026 release, sum of
adjacent-year splits from current prices in $t-1$ to previous-year
prices in $t$; the sample is the economies whose volume service share
rose by more than $.02$. These are accounting allocations of observed
changes, not estimates of causal mechanisms.
\end{tablenotes}
\end{threeparttable}
\end{table}

The production block is split as follows. Write the domestic
coefficient matrix as $A=T\circ d$. Here $T=A+A^{M}$ holds total
(domestic plus imported) input coefficients, $d$ is the domestic share
of each input, and $\circ$ is the elementwise product. A falling
$d_{ij}$ means that imports replace domestic supply of input $i$ to
industry $j$. The block's Shapley value averages its effect over the 32
coalitions of the other five components. For each coalition, I move
$(T,d,v)$ along the straight line from their 1995 values to their 2018
values. Each part receives the integral of the change in $s^S$ that its
own movement causes. This is the Aumann--Shapley value
\citep{aumann1974values}. The parts sum to $\psi^{(A,v)}$ in every
economy, up to a quadrature error below $10^{-15}$, and no order of the
parts enters. Every point on the line is a feasible production block:
no column gets a negative residual $m_j$ that its two endpoints do not
already have. A discrete Shapley split among the same nine parts would
evaluate infeasible blocks in every economy, between 142 and 508 of its
512 synthetic blocks. Its allocations are still almost the same as the
path split: the correlation across the 76 economies is above $.999$ for
every part. The chain-volume version sums the splits of adjacent-year
transitions from current prices in $t-1$ to previous-year prices in
$t$, as in Section~\ref{sec:prices}.

\section{Deflator Audit}
\label{app:deflators}

The previous-year-price tables can be checked against national
accounts. Each pair of current-price and previous-year-price tables
gives a chain-linked value-added deflator for any group of industries.
I compare the price of services relative to industry over 1995--2018
with the same ratio in the World Development Indicators. The
definitions match: services are ISIC G--U and industry is B--F. The two agree
closely. Across the 70 economies with national data, the changes
correlate 0.90. For OECD members the median gap is 0.03 log points.

A rule fixed before computing the gaps flags 8 of 76 economies. An
economy is flagged if its relative-price change differs from the
national accounts by more than 0.25 log points. It is also flagged if
its real value-added growth differs from real GDP growth by more than
0.20 log points. A relative-price error of 0.25 moves a two-sector real
share near one half by about 6 points. The flagged economies are
Brunei, Chile, China, C\^ote d'Ivoire, Morocco, Pakistan, Russia, and
Vietnam. China has the largest gap. In the tables its service price
rises about fivefold relative to industry; in the national accounts it
rises 81 percent. Two features produce the gap. Services' gross-output
prices rise by a factor of 2.9 relative to manufacturing. And double
deflation lowers manufacturing's value-added deflator to 0.44 of its
output deflator, because the prices of manufacturing inputs rose faster
than the prices of its output. The same ratio is 0.96 for India and
0.76 for the United States. Take the two-sector measure, services over
services plus industry. The World Development Indicators split China's
15.9-point nominal rise into 1.1 points of real change and 14.8 points
of relative-price change (Appendix~\ref{app:construction}). The
inter-country tables instead give a real two-sector share that falls
about 25 points at either reference year. With single deflation,
China's market-service share rises 1.4 points over 1995--2018 in volume
instead of falling 6.2.

Dropping the flagged economies changes no conclusion
(Table~\ref{tab:deflators}). $R$ falls in 49 of the 58 remaining
economies whose volume service share rose, 84 percent. With both
changes above two points, the count is 38 of 54. The fall in the median
$R$ is 75 percent complete by 2008, as in the full sample. Single
deflation, which deflates value added by each industry's output price,
gives 55 of 65. Across the price bases and samples in the table, the
count stays between 80 and 86 percent.

Most of the gap between volume and current-price changes in $R$ does
not come from industry prices. Suppose each industry's value added is
deflated by one price for all its uses. Then the volume and
current-price changes in $R$ correlate 0.97, and $R$ changes sign in
one economy instead of nine. The rest of the gap comes from prices that
differ across uses. The OECD estimates these prices by RAS from
deflated final-demand totals, and national accounts do not publish
them. They drive the largest gaps. In Nigeria, the price of market
services sold to households rises about sixtyfold relative to the price
of market-service exports over 1995--2018. That is why its $R$ rises 23
points at current prices but falls 1.5 points in volume. Belarus and
Cambodia show the same pattern in the other direction.

Two caveats apply. First, six economies cannot be tested on relative
prices: C\^ote d'Ivoire, Laos, Myanmar, Tunisia, Taiwan, and the United
States lack sectoral constant-price series that start by 2000. Second,
the income share of the fall is sensitive to the sample in volume. It
ranges from 0.25 to 0.46 across the samples in
Table~\ref{tab:deflators}, inside the interval reported in
Table~\ref{tab:period}.

\begin{table}[H]
\centering
\begin{threeparttable}
\caption{Chain-volume results by price basis and sample, 1995--2018}
\label{tab:deflators}
\scriptsize
\setlength{\tabcolsep}{4pt}
\begin{tabular}{lccccc}
\toprule
 & & $R$ fell, & Both changes & Share of the & Income share \\
Price basis and sample & Economies & $s^S$ rose & above 2 points & fall by 2008 & of the fall \\
\midrule
Current prices, all economies & 76 & 50/65 (77\%) & 42/56 & 0.69 & 0.49 \\
Chain volume, all economies & 76 & 54/64 (84\%) & 40/59 & 0.75 & 0.38 \\
\quad without the 8 flagged economies & 68 & 49/58 (84\%) & 38/54 & 0.75 & 0.39 \\
\quad without flagged and untestable & 63 & 48/56 (86\%) & 37/52 & 0.75 & 0.25 \\
\quad stricter relative-price rule (0.15) & 61 & 44/52 (85\%) & 35/48 & 0.74 & 0.35 \\
Single deflation, all economies & 76 & 55/65 (85\%) & 40/60 & 0.76 & 0.38 \\
\quad without the 8 flagged economies & 68 & 49/58 (84\%) & 38/55 & 0.76 & 0.39 \\
One deflator per industry, all economies & 76 & 51/64 (80\%) & 41/59 & 0.71 & 0.46 \\
\bottomrule
\end{tabular}
\begin{tablenotes}\footnotesize
\item ``$R$ fell, $s^S$ rose'': economies whose service share rose, and
among them those whose $R$ fell. ``Both changes above 2 points'':
$\Delta s^S>.02$ and $\Delta R<-.02$. Flagged economies fail a rule
fixed in advance: the change in the relative price of services to
industry, or real value-added growth, differs from the World
Development Indicators by more than 0.25 or 0.20 log points. Untestable
economies lack the national series. Single deflation deflates each
industry's value added by its output price. One deflator per industry
applies the industry's value-added deflator to all its uses. The income
share is the part of the mean fall in $\log R$ that follows income
growth, as in Table~\ref{tab:period}. \textit{Source}: OECD
inter-country input-output tables in current and previous-year prices,
2026 release; World Development Indicators.
\end{tablenotes}
\end{threeparttable}
\end{table}

\section{A Composition Condition for Servicification}
\label{app:composition}\label{sec:composition}

Section~\ref{sec:accounting} notes that the production block's
contribution to $\Delta R$ has mixed signs across economies. This
heterogeneity reflects composition. The following proposition gives a condition that can be evaluated in
the data.
Collapse final demand into goods and services. Let
$q_g=e_S'\widehat v L u_g$ and $q_s=e_S'\widehat v L u_s$ denote the
service value added embodied in one unit of final delivery of goods and
services, respectively, with $u_g$ and $u_s$ as the corresponding unit
selectors. With household final demand $(f_g^C,f_s^C)$ and
total final demand $(f_g,f_s)$, the general definition
$R=V_C^S/V^S$ becomes
\[
\begin{split}
R
&= \frac{q_g f_g^C+q_s f_s^C}{q_g f_g+q_s f_s}
 = R(a)
 = \frac{f_s^C+a\,f_g^C}{f_s+a\,f_g},
\qquad a\equiv\frac{q_g}{q_s},\\
\frac{\partial R}{\partial a}
&= \frac{f_g^C f_s-f_s^C f_g}{(f_s+a f_g)^2},
\qquad
\operatorname{sign}\Bigl(\frac{\partial R}{\partial a}\Bigr)
= \operatorname{sign}\Bigl(\frac{f_g^C}{f_g} - \frac{f_s^C}{f_s}\Bigr),
\end{split}
\]
holding final-demand structure fixed. So $a$ is not a raw direct
input coefficient: it is the embodied service-value-added intensity
of goods relative to that of services. An increase in $a$ gives more
weight to the goods-final-demand channel, whose household exposure is
$f_g^C/f_g$, relative to the service-final-demand channel, whose exposure
is $f_s^C/f_s$. It raises the domestic household share only where households'
share of goods final demand exceeds their share of service final
demand.

The condition extends to the 50-industry system.

\begin{prop}[Multisector composition condition]
\label{prop:composition}
For any small perturbation $(dA,dv)$ of the production structure, with
final demand held fixed, the first-order response of the domestic
household share is
\[
dR \;=\; \frac{1}{V^S}\sum_j \delta_j\, x_j\,
\bigl(\omega^C_j - R\bigr),
\qquad
\delta_j \;=\; \sum_i q_i\, dA_{ij} \;+\;
\mathbf{1}\{j \in \mathcal S\}\, dv_j,
\]
where $x_j$ is gross output, $\omega^C_j$ the household-supported share
of industry $j$'s output, and $V^S$ service value added. Here
$\mathbf{1}\{\cdot\}$ is the indicator function and $q_i$ the service
value added embodied in a unit delivery of industry $i$. To first
order, the perturbation raises $R$ if and only if
$\sum_j \delta_j x_j(\omega^C_j - R) > 0$.
\end{prop}

\noindent\textit{Proof.}
Write $R=V_C^S/V^S$ with $V_C^S=e_S'\widehat v x^C$ and
$V^S=e_S'\widehat v x$, where $x^C=Lf^C$, $x=Lf$,
$\widehat v=\operatorname{diag}(v)$, and $L=(I_n-A)^{-1}$. For a
perturbation $(dA,dv)$ with final demand fixed, let
$d\widehat v=\operatorname{diag}(dv)$. Since $dL=L\,(dA)\,L$,
\[
dV_C^S=e_S'(d\widehat v)x^C+e_S'\widehat v L\,(dA)x^C .
\]
The row vector $q'=e_S'\widehat v L$ prices a unit delivery of industry
$i$ by its embodied service value added. So the second term is
$\sum_j \bigl(\sum_i q_i\, dA_{ij}\bigr) x^C_j$, and the selector
$e_S'$ confines the first term to service industries,
$\sum_{j\in\mathcal S} dv_j\, x^C_j$. Adding the two terms gives
$dV_C^S = \sum_j \delta_j x^C_j$ with $\delta_j$ as stated. The
identical calculation for total service value added gives
$dV^S = \sum_j \delta_j x_j$. It follows that
\[
dR = \frac{dV_C^S - R\,dV^S}{V^S}
   = \frac{1}{V^S}\sum_j \delta_j\, x_j
     \bigl(\omega^C_j - R\bigr),
\qquad \omega^C_j = x^C_j / x_j . \qquad\blacksquare
\]

The two-sector condition above is the special case with one goods and
one service industry. In efficient economies, Hulten-type results price
such perturbations by Domar weights
\citep{Hulten1978,baqaee2019macroeconomic}; the condition here prices
nothing and is accounting on measured tables, stating only how the two
shares move. Here $d$ is the differential operator, and domestic status
is already part of the definition of $A$. The formula holds for any
perturbation of $A$ and $v$; it does not need the column identity
$v_j+\sum_i A_{ij}+m_j=1$. Because it is exact for the differential,
its sign condition is an exact \textit{local} condition. When the sum
is close to zero, second-order terms can decide the sign of a finite
change. So correctly signing a finite 1995-to-2018 switch, as below, is
an approximation result, not a finite-change theorem.

The evaluation uses a servicification-only switch. For each economy,
the market-service input rows of the input-coefficient matrix move from
their 1995 to their 2018 values, column by column. Each column's
value-added coefficient is displaced by the added service requirement,
so the accounting identity holds; 24 of 3,800 columns would violate
nonnegativity and are left unswitched. All final demand is held at
1995. The first-order condition correctly signs the discrete response
in all 76 economies on the 2025-release tables, including both positive
and negative responses. China is a negative case (first-order
prediction $-0.0282$, realized $-0.0233$); Ireland a positive one
($+0.0281$, $+0.0391$); and Korea negative ($-0.0047$, $-0.0045$). The
exercise evaluates the condition on measured input-output structures;
it does not test a behavioral mechanism, because the outcome is
generated by the counterfactual switch the condition summarizes. Still,
it clarifies the interpretation of Table~\ref{tab:shapley}'s mixed
production-block signs. Whether greater use of service inputs raises or
lowers $R$ is predicted, economy by economy, by who ultimately buys the
output of the servicifying industries.

\section{Accounting Decomposition Details}
\label{app:accounting}

Table~\ref{tab:mechanismmoments} reports the Shapley shares of
Section~\ref{sec:accounting} by income tercile.

\begin{table}[H]
\centering
\begin{threeparttable}
\caption{Shapley shares by income tercile among economies with
material service-share increases, 1995--2018}
\label{tab:mechanismmoments}
\scriptsize
\setlength{\tabcolsep}{4pt}
\begin{tabular}{lrrrr}
\toprule
& &\multicolumn{2}{c}{Median contribution share}&\\
\cmidrule(lr){3-4}
2018 income tercile&$N$&Production system&
Household composition&Fraction PS largest\\
\midrule
Low&15&.57&.50&.60\\
Middle&21&.67&.27&.81\\
High&20&.91&.13&.95\\
\midrule
All&56&.73&.26&.80\\
\bottomrule
\end{tabular}
\begin{tablenotes}
\footnotesize
\item Economies with $\Delta s^S>.02$. Income groups are terciles of
2018 GDP per capita (PPP) among the 76 economies. The middle two
columns report medians of the within-economy Shapley contribution
divided by $\Delta s^S$. ``Production system'' combines $(A,v)$,
$b^I$, and $b^X$. ``Largest'' compares that combined contribution
with the three remaining exhaustive groups: $b^C$, $b^G$, and
$\varphi$. Because channels can offset one another, contribution
shares need not lie in $[0,1]$ and row medians need not sum to one.
Using 2022 instead of 2018 as the endpoint leaves the production
system largest in 40 of 51 material tertiarizers (78 percent), with a
median share of .96 in the high-income tercile.
\end{tablenotes}
\end{threeparttable}
\end{table}

The Shapley accounting components use the same OECD domestic-use tables as the
measurement sections. $A$ is domestic intermediate use divided by gross
output; $v$ is value added divided by gross output; and the residual
$m=\mathbf{1}-v-\mathbf{1}'A$ contains imported intermediate inputs and net taxes.
Final-demand vectors are household consumption (HFCE), government
(NPISH plus government consumption), investment (GFCF plus inventories),
and exports (exports plus non-resident purchases). Each category vector
is separated into its total-final-demand share $\varphi_k$ and its
industry composition $b^k$. The baseline uses 1995 and 2018 because
both years have broad country coverage and 2018 is the last
pre-pandemic benchmark year.

The Shapley decomposition comes from cooperative game theory, but it
is an established tool in empirical economics.
\citet{shorrocks2013decomposition} develops it as the unified framework
for decomposing distributional statistics into contributing factors.
It is routinely applied to inequality accounting by factor components
\citep{sastre2002shapley} and, in development settings, to poverty
decompositions \citep{kolenikov2005regional}.
The use here is the same exercise applied to input-output statistics: an
exact, order-free attribution of the change in a published aggregate
to the measured components that enter it.

The main allocations hold across samples and cutoffs. On the
absolute-change sample ($|\Delta s^S| > .02$), the median combined
production-system share is $.73$ at material-change cutoffs of $.01$,
$.02$, and $.03$ alike. The value-weighted aggregate is $.74$, and the
10th--90th percentile range is $[.30, 1.22]$. Alternative groupings
move the share sensibly: adding $\varphi$ to the group gives $.77$, and
the production block alone $.47$. Restricting to rising
shares ($\Delta s^S > .02$) leaves the median at $.73$. On the $R$
side, the $\varphi$ share is $.97$ on the absolute-change sample and
$1.00$ restricted to material declines, with a negative $\varphi$
contribution in every declining economy.

For the feasibility audit, every mixed-year pair $(A_{t_a},v_{t_v})$ is
evaluated against $m=\mathbf{1}-v-\mathbf{1}'A$. A pair is infeasible if
any industry has $m_j<0$. Ninety percent of mixed-year combinations
violate this restriction in at least one column. Of the 152 measured
endpoint blocks, 151 satisfy $m_j\geq0$ in every column. The exception
is Romania's 1995 food-products column (C10T12), whose implied residual
is $-0.000174$, a 0.017 percentage-point discrepancy attributable to
rounding in the published table. I retain the published $A$ and $v$
coefficients without renormalization. The grouped Shapley switches
$(A,v)$ jointly and so never combines their values from different
years. Exact
enumeration evaluates all $2^6$ coalitions and computes the standard
factorial Shapley weights; the maximum adding-up residual is numerical
rounding.

The median economy is not the aggregate. As a descriptive
aggregation, the national flows of the 76 economies can be pooled in
current U.S. dollars. This produces a size-weighted aggregate of the
covered national accounts, not a consolidated world input--output
system with bilateral transactions. In the pooled aggregate, the
service share rises from 0.498 to 0.513 and the domestic household
share falls from 0.595 to 0.534 between 1995 and 2018. The same grouped
decomposition assigns $+0.0136$ of the $+0.0148$ pooled service-share
change to the production block, roughly nine-tenths, against $+0.0014$
(under a tenth) for household composition. It assigns $-0.0575$ of the
$-0.0611$ pooled decline in $R$ to the final-demand structure, or 94
percent. These pooled results place greater weight on large economies
and should be read separately from the median-country decomposition.

\bibliographystyle{aer_nodash}
\bibliography{references}

\end{document}